\documentclass[aps,pra,showpacs,twoside,twocolumn,10pt,floatfix,nofootinbib,longbibliography]{revtex4-1}
\usepackage[colorlinks=true, citecolor=red, urlcolor=blue ]{hyperref}
\usepackage{epsfig,newlfont,amssymb,amsfonts,amsmath,bm,subfigure,palatino,mathtools,amsthm,braket,soul,enumitem,color,times,comment,geometry,graphics}
\usepackage[normalem]{ulem}
\begin{document}
\title{Distinguishing phases via non-markovian dynamics of entanglement in topological quantum codes under parallel magnetic field}
\author{Harikrishnan K. J. and Amit Kumar Pal}
\affiliation{Department of Physics, Indian Institute of Technology Palakkad, Palakkad 678 623, India}
\date{\today}

\begin{abstract}
We investigate the static and the dynamical behavior of localizable entanglement and its lower bounds on nontrivial loops of topological quantum codes with parallel magnetic field. Exploiting the connection between the stabilizer states and graph states in the absence of the parallel field and external noise, we identify a specific measurement basis, referred to as the canonical measurement basis, that optimizes localizable entanglement when measurement is restricted to single-qubit Pauli measurements only, thereby providing a lower bound. In situations where computing even the lower bound is difficult, we propose an approximation of the lower bound that can be computed for larger systems according to the computational resource in hand. Additionally, we also compute a lower bound of the localizable entanglement that can be computed by determining the expectation value of an appropriately designed witness operator. We investigate the behavior of these lower bounds in the vicinity of the topological to nontopological quantum phase transition of the system, and perform a finite-size scaling analysis. We also investigate the dynamical features of these lower bounds when the system is subjected to Markovian or non-Markovian single-qubit dephasing noise. We find that in the case of the non-Markovian dephasing noise, at large time, the canonical measurement-based lower bound oscillates with a larger amplitude when the initial state of the system undergoing dephasing dynamics is chosen from the nontopological phase, compared to the same for an initial state from the topological phase. On the other hand, repetitive collapses followed by revivals to high value with time is observed for the proposed witness-based lower bound in the nontopological phase, which is absent in the topological phase. These features can be utilized to distinguish the topological phase of the system from the nontopological phase in the presence of dephasing noise.
\end{abstract}

\maketitle

\section{Introduction}
\label{sec:intro}

The world-wide drive for achieving quantum supremacy~\cite{preskill2012,harrow2017,arute2019} and for implementing large-scale fault-tolerant quantum computers~\cite{divincenzo2000,gottesman2010} in the last couple of decades have established topological quantum error correcting codes~\cite{fujii2015,nayak2008,pachos2014,lahtinen2017}, eg. the Kitaev code~\cite{kitaev2001,kitaev2003,kitaev2006} and the color code~\cite{bombin2006,bombin2007} as ideal candidate systems for the task. Robustness of these systems against loss of physical qubits~\cite{stace2009,stace2010,vodola2018,stricker2020}, and computational errors~\cite{nayak2008,pachos2014,lahtinen2017,wootton2011,katzgraber2009} has motivated realizations of these systems in the laboratory using substrates like trapped ions~\cite{nigg2014,linke2017,wright2019} and superconducting qubits~\cite{kelly2015,gambetta2017}, which has made experimental verification of theoretical results possible. Moreover, with the introduction of noisy intermediate-scale quantum devices built using $50$-$100$ physical qubits~\cite{preskill2018,paler2018,nash2019}, and their use towards the goal of achieving quantum supremacy~\cite{arute2019}, the importance of topological quantum codes hosting a large number of qubits has now been established in the context of building large quantum memories~\cite{kitaev2001} and successfully implementing quantum error correction protocols for errors on multiple physical qubits~\cite{fujii2015,nayak2008,pachos2014,lahtinen2017}. 

The possibility of adverse effects of local perturbations in quantum computation tasks has motivated investigations of topological quantum codes as lattice models~\cite{wootton2011,trebst2007,wu2012,dusuel2011,zarei2019,tsomokos2011,wiedmann2020,karimipour2013,jamadagni2018,zarei2015,jahromi2013,zarei2015}, where perturbations in the form of external magnetic fields~\cite{wootton2011,trebst2007,wu2012,dusuel2011,zarei2019,tsomokos2011,zarei2015,jahromi2013} and spin-spin interactions~\cite{wiedmann2020,karimipour2013,zarei2015} are considered. The ground state of the system retains the topological order~\cite{laughlin1983,wen1990,wen1995} when the external perturbation is small, while with increasing perturbation strength, a topological to nontopological quantum phase transition (QPT) takes place, the QPT point being a quantifier of the robustness of the code corresponding to the perturbation parameter. In contrast to the Landau paradigm of descriptions of phases and order parameters~\cite{landau1937,goldenfeld1992}, topological to nontopological phase transitions cannot be characterized by local order parameters and spontaneous symmetry breaking~\cite{wen1990,wen1995}. While studies of the topological to nontopological QPTs in locally perturbed topological quantum codes has so far been carried out in terms of ground state energy per site and the single-particle gap~\cite{wootton2011,trebst2007,wu2012,dusuel2011,zarei2019,tsomokos2011,wiedmann2020,karimipour2013,jamadagni2018,zarei2015,jahromi2013,zarei2015}, the importance of the sustenance of the topological order against local perturbations in the quantum information processing and quantum computation~\cite{fujii2015,nayak2008,pachos2014,lahtinen2017} highlights the necessity of investigations of these systems in terms of quantum correlation measures that are resources in quantum protocols~\cite{horodecki2009,modi2012,bera2017}.  

Along with serving as resource in quantum protocols like teleportation~\cite{horodecki2009,bennett1993,bouwmeester1997}, dense-coding~\cite{horodecki2009,bennett1992,mattle1996,sende2011}, quantum cryptography~\cite{horodecki2009,ekert1991,jennewein2000}, and measurement-based quantum computation~\cite{horodecki2009,raussendorf2001,raussendorf2003,briegel2009}, entanglement~\cite{horodecki2009,guhne2009} is by far the most widely accepted quantum correlation for characterizing quantum many-body systems~\cite{amico2008,dechiara2018}, including nontopological~\cite{amico2008,dechiara2018,skrovseth2009,smacchia2011,montes2012} and topological phases~\cite{kitaev2003,kitaev2006,kargarian2008} of lattice models. Advancement in the investigation of biparty- and multiparty-entangled quantum states in the laboratory using trapped ions~\cite{leibfried2003,leibfried2005,brown2016} and superconducting qubits~\cite{clarke2008,berkley2003} has brought testing of theoretical results on entanglement in characterizing phases of perturbed topological quantum codes within our grasp. In this line of investigation, a number of challenges have been prominent. The fact that topological phases are characterized by non-local order parameters~\cite{laughlin1983,wen1990,wen1995} indicates the requirement of investigating multipartite entanglement~\cite{horodecki2009} in the ground state(s) of the system, or in the reduced state of a chosen subsystem, which is difficult due to the scarcity of computable multiparty entanglement measures~\cite{horodecki2009}. Also, partial trace-based approach of computing entanglement over a chosen subsystem of topological quantum codes by determining its reduced state from the ground state of the entire system fails as tracing out the degrees of freedom of the spins in the rest of the system results in diagonal density matrices, thereby leading to vanishing entanglement~\cite{horodecki2009,amaro2018,amaro2020a}. Moreover, inevitable interaction of the system with environment~\cite{breuer2002,rivas2012,rivas2014} leads to a rapid decay of entanglement over time~\cite{horodecki2001,diosi2003,dodd2004,almeida2007,salles2008,yu2009}, making it difficult to investigate the phase structure of the perturbed topological quantum code in terms of entanglement at a latter time in the realistic scenario. Although the effect of thermal noise on entanglement in unperturbed topological quantum codes has been investigated~\cite{castelnovo2007,castelnovo2008,schmitz2019}, trends of entanglement in these systems in the presence of noise as well as local perturbations in the form of magnetic field or spin-spin interaction remain unexplored.  

\begin{figure*}
    \centering
    \includegraphics[width=0.8\textwidth]{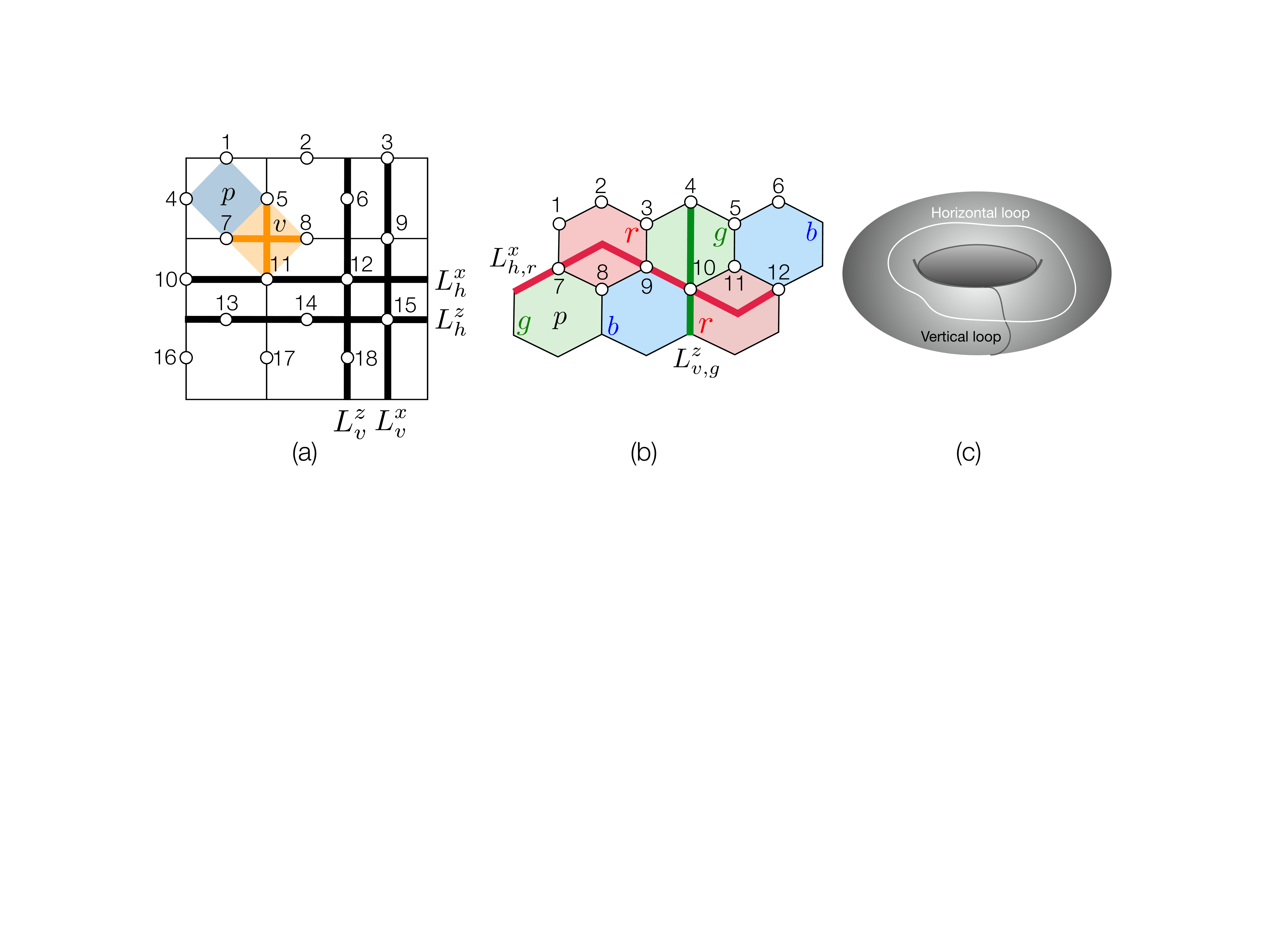}
    \caption{(Color online) \textbf{(a) A Kitaev code on a square lattice of $3\times 3$ architecture}. Each of the $N_E$ edges in the square lattice of $N_P$ plaquettes contains a qubit, represented by the circles, constituting an $N$-qubit ($N=N_E=2N_P=18$ in the present case, with $N_p^h=N_P^v=3$) system with periodic boundary conditions in both horizontal and vertical directions. Here, $N_E$ $(N_P)$ is the cardinality of $\mathcal{N}_E$ ($\mathcal{N}_P$) (see Section~\ref{sec:definitions}). A plaquette, denoted by the index $p$ (shown by a shaded square),  and a vertex, denoted by the index $v$ (shown by thick intersecting lines on a shaded square) consist four qubits each. The nontrivial loops in the vertical and horizontal direction, representing respectively the logical operators $L^z_{h,v}$ and $L^x_{h,v}$, have been marked with the thick lines consisting of three qubits each. The $L_h^x$ ($L_{h}^z$) and the $L_v^z$ ($L^x_h$) operators intersect each other at a single qubit. Periodic boundary condition is assumed in both horizontal and vertical directions. \textbf{(b) Color code on a hexagonal lattice}. The lattice is constituted of $N_P$ plaquettes (in the present case, $N_P=6$ in a $3\times 2$ architecture) and $N$ (in this example, $N=12$) qubits. Red ($=c_1=r$), green ($=c_2=g$), and blue ($=c_3=b$) signify three different colors of the plaquettes. The thick vertical and horizontal lines represent respectively the vertical and horizontal nontrivial loops representing the operators $L^\alpha_{h,r}$ and $L^\beta_{v,g}$ ($\alpha=x(z)$ if $\beta=z(x)$). Periodic boundary condition is assumed in both horizontal and vertical directions.  \textbf{(c) A torus of genus $1$.} The horizontal and the vertical nontrivial loops are shown by the white (horizontal) and gray (vertical) lines.}
    \label{fig:fig1}
\end{figure*}

In this paper, we address the question as to whether the topological and nontopological  phases and the corresponding QPT can be investigated in terms of appropriate entanglement measures, both in the absence as well as presence of decoherence in the system. We consider topological quantum codes, such as the Kitaev code and the color code, in the presence of a parallel magnetic field~\cite{trebst2007,wu2012,zarei2015,jahromi2013}, and quantify entanglement over a multiparty subsystem of the code via a local measurement-based protocol~\cite{divincenzo1998,verstraete2004,popp2005,verstraete2004a,jin2004,sadhukhan2017}, where a \emph{localizable entanglement}~\cite{verstraete2004,popp2005} can be computed over a chosen subsystem by maximizing the average entanglement over the subsystem with respect to all possible single-qubit projection measurements performed on all the qubits in the rest of the system. The choice of such a measure of entanglement is based on the recent results on the multiparty nature of localizable entanglement~\cite{banerjee2020}, and the requirement for a non-local order parameter to characterize the topological phase of the system.  We show that apart from being useful in introducing concepts like correlation length in low-dimensional quantum spin models~\cite{verstraete2004a,jin2004}, characterizing phases in cluster-Ising~\cite{skrovseth2009,smacchia2011} and cluster-XY models~\cite{montes2012}, and as the key resource in protocols like measurement-based quantum computation~\cite{raussendorf2001,raussendorf2003,briegel2009} and entanglement percolation\cite{acin2007}, localizable entanglement can also aid in investigating topological to nontopological quantum phase transitions occurring in topological quantum codes under parallel magnetic field, and also under single-qubit dephasing noise.  

To tackle the difficulty of computing the localizable entanglement over a chosen subsystem due to the measurement-based optimization involved in its definition~\cite{verstraete2004,popp2005,verstraete2004a,jin2004},  we numerically compute a number of lower bounds in topological quantum codes of increasing sizes (cf.~\cite{amaro2018,amaro2020a}), as a function of the strength of the parallel magnetic field, over nontrivial loops on the lattice under periodic boundary condition.  When the external field strength is zero, the ground states of the topological codes can be connected to graph states~\cite{hein2006} via local Clifford operations~\cite{lang2012,amaro2018,amaro2020a,HK_inprep}. Using this, we identify a canonical setup of Pauli measurements which optimizes the lower bound of localizable entanglement when measurements are restricted to single-qubit Pauli measurements. In the case of the Kitaev code, the canonical measurement setup can be described using the positions of the qubits in the lattice relative to the plaquetes and vertex stabilizers through which the nontrivial loop passes. In the presence of the external parallel field, the canonical measurement setup provides a lower bound of the localizable entanglement. In situations where the computation of even the canonical measurement-based lower bound proves difficult due to the large size of the system, we propose an approximation of the lower bound that can be determined depending on the computational resource in hand. We demonstrate that this approximation provides the lower bound with negligible error in the case of Kitaev code of large size under parallel magnetic field. We also consider a lower-bound of localizable entanglement that can be computed using the expectation value of an appropriately designed witness operator~\cite{guhne2009,eisert2007,guhne2007} for the nontrivial loop in the absence of the parallel magnetic field. The witness operator can be constructed in terms of the stabilizer operators of the code obeying a specific set of rules~\cite{guhne2005,alba2010,amaro2020}, and has a one-to-one correspondence of the chosen canonical measurement setup~\cite{amaro2018,amaro2020a}. This provides an avenue to experimentally probe the results described in this paper.

We investigate the behavior of the canonical measurement-based and witness-based lower bounds of localizable entanglement across the topological to nontopological quantum phase transition that the system undergoes when the strength of the parallel magnetic field is increased. We demonstrate that in the case of the Kitaev code, the absolute value of the first derivative of both the bounds with respect to the field strength exhibits a maximum in the vicinity of the quantum phase transition point. We also perform a finite-size scaling analysis corresponding to the approach of the position of the maximum towards the quantum phase transition point with increasing system size. We find that although the performance of the witness-based lower bound diminishes with the introduction of the parallel magnetic field, the behaviors of it's first derivative remains unchanged across the quantum phase transition point. We also find that the finite-size effect is more prominent in the case of localizing entanglement over nontrivial loops corresponding to the logical $Z$ operators, compared to the same for logical $X$ operators. Although the investigation of the quantum phase transition in the color code becomes difficult due to the rapid increase in the system size, our results regarding small color code indicate that similar behavior of the localizable entanglement across the topological to nontopological quantum phase transition in the color codes with parallel magnetic field can be expected.       

We also assume a situation where each of the qubits in the system is subjected to local Markovian and non-Markovian dephasing noise~\cite{luczka1990,palma1996,reina2002,haikka2011,haikka2013}, and ask if the topological and nontopological phases can be distinguished during the evolution of the system. To answer this question, we look into the dynamics of the canonical measurement-based and witness-based lower bounds of localizable entanglement. We show that in the presence of non-Markovian single-qubit dephasing noise on all qubits due to their connection with local baths constituted of simple harmonic oscillators with Ohmic spectral function~\cite{haikka2011,haikka2013}, at a large time, the canonical measurement-based lower bound of localizable entanglement exhibits rapid oscillation with high amplitude as a function of the strength of the parallel magnetic field in the nontopological phase. This is in contrast to the behavior of the lower bound of localizable entanglement in the topological phase exhibiting the absence of an oscillation or an oscillation with low amplitude, thereby distinguishing the two phases. The oscillation increases with an increase in the value of the Ohmicity parameter corresponding to the dephasing noise. On the other hand, repetitive collapses followed by revivals to high value of the witness-based lower bound is found  during its dynamics, when the initial state of the dynamics under non-Markovian noise is chosen from the nontopological phase, in contrast to the absence of such behavior in topological phase. These features can be used to distinguish between the phases of the model even when the system is undergoing evolution under dephasing noise. 

The paper is organized as follows. In Section~\ref{subsec:system}, we provide brief descriptions of the topological quantum codes, including the Kitaev code and the color code, in the presence of parallel magnetic field. Section~\ref{subsec:noise} describes single-qubit dephasing noise and the corresponding quantum master equation. The definitions of localizable entanglement and its lower bounds, including the witness-based lower bound, are provided in Section~\ref{subsec:entanglement}. The static properties of the localizable entanglement in the topological codes under parallel magnetic field are discussed in Section~\ref{sec:optimal_basis}. In Section~\ref{subsec:canonical}, we discuss the connection between the stabilizer ground states of the topological codes and the graph states, and introduce the canonical measurement setup. We also present the approximation scheme for the canonical measurement-based lower bound, estimate the error in this approximation, and demonstrate its efficiency in the case of large systems. We also discuss the behavior of the canonical measurement-based lower bound across the topological to nontopological quantum phase transition in the Kitaev model with the increasing magnetic field strength, and perform the finite-size scaling analysis. Similar analysis is carried out for the witness-based lower bound in Section~\ref{subsec:qpt_witness}. We also discuss our results in the context of color codes under parallel magnetic field. Section~\ref{sec:dynamics} describes the behavior of the lower bounds of localizable entanglement as a function of time, when the system is subjected to Markovian and non-Markovian single-qubit dephasing noise. Distinguishing between the topological and the nontopological phases of the model using the large time dynamics of the lower bounds of localizable entanglement is discussed in Section~\ref{subsec:distinction}. Section~\ref{sec:conclusion} contains the concluding remarks, and a discussion on possible future directions.

\section{Models and Methodology}
\label{sec:definitions}

In this Section, we provide a brief overview of the topological codes in the presence of parallel magnetic field. We also discuss the Markovian and non-Markovian dephasing noise, and define localizable entanglement as the appropriate measure for entanglement in topological quantum codes.  

\subsection{Topological codes in parallel magnetic field}
\label{subsec:system}

We start our discussion with an overview of the topological quantum codes investigated in this paper. 

\subsubsection{Kitaev code in a parallel magnetic field} 
\label{subsubsec:kitaev}

Let us consider a 2D rectangular lattice where the sets of all edges and plaquettes are denoted by $\mathcal{N}_E$ and $\mathcal{N}_P$ respectively, hosting a total of $N$ qubits represented by spin-$\frac{1}{2}$ particles. The lattice is constructed on a $N_P^h\times N_P^v$ architecture, where $N_P^h$ $(N_P^v)$ is the number of plaquettes in the horizontal (vertical) direction, such that $N_P=N_P^hN_P^v$ is the total number of plaquettes in the system (see Figure~\ref{fig:fig1}(a)). Each qubit in the system is situated on an edge of the lattice. Two types of stabilizer operators -- the plaquette operators $S_p$ and the vertex operators $S_v$ -- are defined on this lattice as 
\begin{eqnarray}
S_p &=& \bigotimes_{i\in p}\sigma^z_i,S_v = \bigotimes_{i\in v}\sigma^x_i, 
\label{eq:kitaev_stabilizers}
\end{eqnarray}
where $\sigma^x_i$ $\left(\sigma^z_i\right)$ is the $x$ $(z)$ component of Pauli matrices, and $p$ $(v)$ is the plaquette (vertex) index. The plaquette (vertex) operator $S_{p}$ $\left(S_v\right)$ has support on the qubits forming the plaquette $p$ (connected directly to the vertex $v$) of the square lattice. The Hamiltonian describing the Kitaev model under a parallel magnetic field is given by~\cite{trebst2007,wu2012}
\begin{eqnarray}
H_K=-J_P\sum_{p}S_p-J_V\sum_{v}S_v-h\sum_{i=1}^N\sigma^z_i,
\label{eq:kitaev_hamiltonian}
\end{eqnarray}
where $J_P$ and $J_V$ are respectively the plaquette and the vertex interaction strengths, and $h$ is the strength of the external magnetic field on every spin. We focus on the parameter subspace defined by $J_P=J_V=J>0$. In the limit $h/J\rightarrow 0$, the Hamiltonian in Eq.~(\ref{eq:kitaev_hamiltonian}) represents the Kitaev model~\cite{kitaev2001,kitaev2003,kitaev2006}.

Under periodic boundary condition, Kitaev model on a square lattice can be embedded on the surface of a torus of genus $1$ (see Figure~\ref{fig:fig1}(c)), which holds two nontrivial loops in horizontal and vertical directions. Let us denote the sets of qubits supporting these nontrivial loops in the horizontal and vertical directions by $\mathcal{S}_h$ and $\mathcal{S}_v$, respectively.  Four nontrivial loop operators of $x$ and $z$ type, corresponding to these loops, can be defined as (see Figure~\ref{fig:fig1}(a))
\begin{eqnarray}
L_\alpha^x=\bigotimes_{i\in\mathcal{S}_\alpha}\sigma_i^x, L_\alpha^z=\bigotimes_{i\in\mathcal{S}_\alpha}\sigma_i^z,
\label{eq:kitaev_loop_operators}
\end{eqnarray}
where $\alpha=h,v$, denoting horizontal and vertical loops, respectively. The Kitaev model has a four-fold degenerate ground state, each of which is an entangled state, 
which can be constructed by applying the nontrivial loop operators on
\begin{eqnarray}
\ket{\psi} &=& \frac{1}{2^N}\prod_v\left[I+S_v\right]|\ket{0}^{\otimes N},
\label{eq:kitaev_ground_state}
\end{eqnarray}
as
\begin{eqnarray} 
\ket{\psi}_g=\left(L_h^x\right)^a\left(L_v^x\right)^b\ket{\psi},
\end{eqnarray}
with $a,b=0,1$, where $\{\ket{0},\ket{1}\}$ is the computational basis in the qubit Hilbert space~\cite{zarei2015}.  With the application of the parallel magnetic field, the ground state degeneracy is lifted, and a non-degenerate ground state of $H_K$ is obtained. The limit $h/J\rightarrow\infty$ corresponds to a fully separable ground state of the form $\ket{0}^{\otimes N}$. With increasing the strength of the parallel magnetic field $h$, there is a topological to nontopological QPT at the critical value $g_c=0.328474(3)$~\cite{trebst2007,wu2012} of the dimensionless system parameter $g=\frac{h}{J}$, which can be determined by identifying the equivalence of the model with the 2D transverse-field Ising model.

\subsubsection{Color code in a parallel field} 
\label{subsubsec:color}

Color codes are topological quantum error correcting codes defined on three-colorable trivalent lattices. The qubits, represented by spin-$\frac{1}{2}$ particles, are situated on the vertices of the lattice~\cite{bombin2006,bombin2007}. On each plaquette, two types of  stabilizer operators are defined as 
\begin{eqnarray}
S^z_p &=& \bigotimes_{i\in p}\sigma^z_i,S^x_p = \bigotimes_{i\in p}\sigma^x_i, 
\label{eq:color_code_stabilizers}
\end{eqnarray}
where $p$ is the plaquette index. The color code Hamiltonian in the presence of a parallel magnetic field on a hexagonal lattice is given by~\cite{jahromi2013}
\begin{eqnarray}
H_C &=& -J\sum_{p}\left(S^x_p+S^z_p\right)-h\sum_{i=1}^N\sigma^x_i,
\label{eq:color_code_hamiltonian}
\end{eqnarray}
where $J$ is the plaquette interaction strength. We consider a hexagonal lattice (see Figure~\ref{fig:fig1}(b)) where the plaquettes can be coded with three different colors such that no two adjacent plaquette has the same color.  Similarly, the lattice links can also be colored with the same three colors such that link of a specific color connects plaquettes of the same color. In this paper, we assume $J>0$. The color code Hamiltonian is obtained from $H_C$ in the $h/J\rightarrow 0$ limit. 

Similar to the square lattice, the hexagonal lattice can also be embedded on the surface of a torus of genus $1$ under periodic boundary condition, and sets of qubits, denoted by $\mathcal{S}_h^c$ and $\mathcal{S}_v^c$ and constituting two nontrivial loops made of links of each of the three colors, can be identified, where $c\in\{c_1,c_2,c_3\}$ denotes the color index. Using these qubits, six fundamental nontrivial loop operators can be constructed as (see Figure~\ref{fig:fig1}(b)) 
\begin{eqnarray}
L_{\alpha,c}^x=\bigotimes_{i\in\mathcal{S}_\alpha^c}\sigma_i^x, L_{\alpha,c}^z=\bigotimes_{i\in\mathcal{S}_\alpha^c}\sigma_i^z.
\label{eq:color_loop_operators}
\end{eqnarray}
Successive applications of these nontrivial loop operators on
\begin{eqnarray}
\ket{\psi} &=& \frac{1}{2^N}\prod_p\left[I+\mathcal{S}_p^x\right]|\ket{0}^{\otimes N},
\label{eq:color_code_groundstate}
\end{eqnarray}
as 
\begin{eqnarray}\ket{\psi}_g=(L_{h,c_1}^x)^{a_1}(L_{h,c_2}^x)^{a_2}(L_{v,c1}^x)^{b_1}(L_{v,c2}^x)^{b_2}\ket{\psi}, 
\end{eqnarray}
$a_1,a_2,b_1,b_2=0,1$ and $c_1,c_2$ being any two of the three colors, generates the ground state manifold $\ket{\psi}_g$ of the $h/J\rightarrow 0$ limit of the Hamiltonian, consisting of 16 degenerate entangled states~\cite{jahromi2013}. On the other hand, a fully polarized ground state is found at $h/J\rightarrow\infty$, with all spins pointing in the field direction. The topological to nontopological QPT occurring with increasing $h/J$ can be determined by mapping the model to a Baxter Wu model in a transverse field on a triangular lattice, where The critical value of the dimensionless system parameter $g=\frac{h}{J}$ is $g_c=0.385$~\cite{jahromi2013}.  

\begin{figure}
    \centering
    \includegraphics[width=0.8\linewidth]{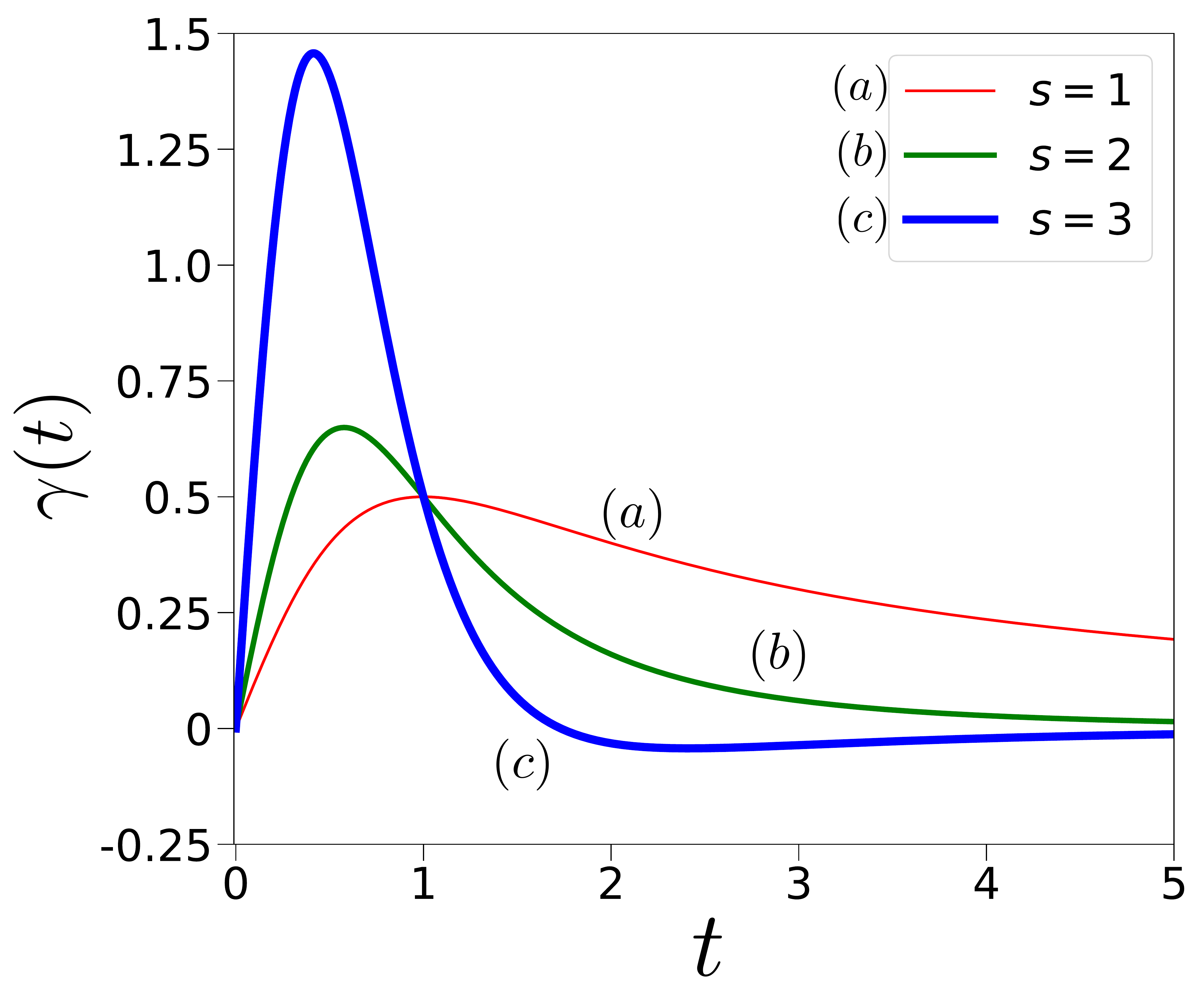}
    \caption{(Color online) \textbf{Dephasing rate.} Variation of $\gamma(t)$ as a function of $t$ for different values of $s$, across the Markovianity to non-Markovianity transition point $s=2$. Both axes are dimensionless.}
    \label{fig:fig2}
\end{figure}

\begin{figure*}
    \centering
    \includegraphics[width=0.7\linewidth]{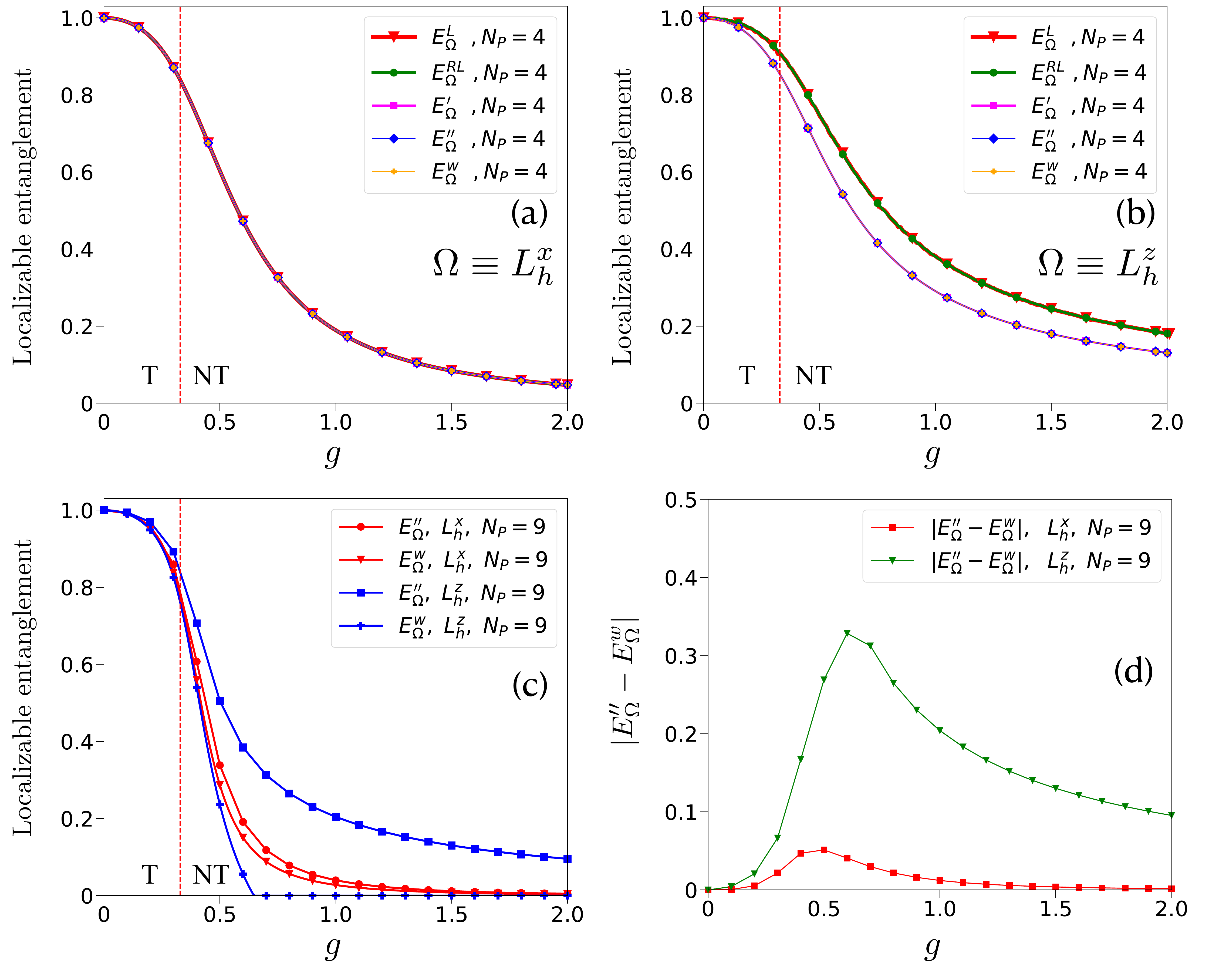}
    \caption{(Color online) \textbf{Bounds of localizable entanglement in Kitaev code.} (a) and (b) Variations of $E^L_\Omega$, $E^{RL}_\Omega$, $E^{\prime}_\Omega$, $E^{\prime\prime}_\Omega$, and $E^w_\Omega$ (vertical axes) as functions of the parallel magnetic field strength $g$ (horizontal axes) in the cases of a Kitaev code of $N_P=4$ plaquettes, where entanglement is localized over a nontrivial loop corresponding to (a) $L^x_{h}$ and (b) $L^z_{h}$. The labels ``T" and ``NT" represent the ``topological" and the ``nontopological" phases on the $g$ axis, and the dashed vertical line indicates the topological to nontopological QPT point $g_c=0.328474(3)$. (c) The dependence of $E^{\prime\prime}_\Omega$ and $E^{w}_\Omega$ (vertical axis) on $g$ (horizontal axis) is shown for a Kitaev code with $N_P=9$ plaquettes, corresponding to the nontrivial loops representing $L^{x}_{h}$ and $L^{z}_h$. (d) The variation of the difference between the two lower bounds, $\left|E^{\prime\prime}_\Omega-E^w_\Omega\right|$ (vertical axis), as a function of $g$ (horizontal axis) in the case of a Kitaev code of $9$ plaquettes, for nontrivial loops representing $L^x_{h}$ and $L^z_h$. Both the horizontal and the vertical axes in all figures are dimensionless.}
    \label{fig:fig3}
\end{figure*}

\subsection{Dynamics under dephasing noise}
\label{subsec:noise}

Let us now consider a situation where each of the spin-$\frac{1}{2}$ particles in the system starts interacting with a bath from a collection of identical and independent thermal baths at time $t=0$. Each bath is made of harmonic oscillators, with a bath Hamiltonian given by $H_b=\sum_k\omega_ka_k^\dagger a_k$, where $\omega_k$ is the frequency of the $k$th bath-mode, and $a^\dagger_k$ $(a_k)$ is the creation (annihilation) operator corresponding to the mode $k$. The interaction between each spin and its bath is given by $H_{sb}=\sum_{k}\sigma^z\otimes\left(g_ka_k+g_k^*a_k^\dagger\right)$, where $g_k$ is the coupling constant between the spin variable and the $k$th mode of the bath, such that in the continuum limit, $\sum_k|g_k|^2$ goes to $\int f(\omega)\delta(\omega_k-\omega)d\omega$, $f(\omega)$ being the spectral function of the bath. Assuming that each spin interacts with its own bath only and is immune to the effects of the remaining baths, and considering thermal initial state of the bath, the time-local quantum master equation that governs the dynamics of the system is given by~\cite{luczka1990,palma1996,reina2002,haikka2011,haikka2013}
\begin{eqnarray}
\dot{\rho}=-\frac{\text{i}}{\hbar}[H,\rho]+\gamma(t)\sum_{i=1}^N\left(\sigma_i^z\rho\sigma_i^z-\rho\right).
\label{eq:qme}
\end{eqnarray}
Here, $H$ is the system Hamiltonian $H_K$ (Eq.~(\ref{eq:kitaev_hamiltonian})) or $H_C$ (Eq.~(\ref{eq:color_code_hamiltonian})), depending on the choice of the system, and $\rho$ is the $N$-qubit state of the system.
The time-dependent dephasing rate $\gamma(t)$ is the same for all qubits, and is given by~\cite{haikka2013}
\begin{eqnarray}
\gamma(t)&=& \omega_c\left[1+(\omega_ct)^2\right]^{-\frac{s}{2}}\sin\left[s\tan^{-1}(\omega_c t)\right]\Gamma(s),
\label{eq:gamma}
\end{eqnarray}
where $\Gamma(s)=\int_0^\infty x^{s-1}\text{e}^{-x}dx$ is the Euler Gamma function,  $\omega_c$ is the cut-off frequency of the bath, and $s$ is the Ohmicity parameter whose value determines whether the bath is sub-Ohmic ($s<1$), Ohmic $(s=1)$, or super-Ohmic ($s>1$) (see Figure~\ref{fig:fig2} for the typical shapes of $\gamma(t)$). At the zero temperature limit of the bath, the value of $s\leq 2$ ensures Markovian spin-bath interaction, while non-Markovianity emerges for $s>2$~\cite{haikka2011,haikka2013}. The critical value $s_c=2$ for the Markovian to non-Markovian transition on the Ohmicity parameter increases with increasing temperature of the bath. In this paper, we focus on the zero temperature limit of the bath, and unless otherwise stated, we fix the bath cut-off frequency to be $\omega_c=1$.

Solving Eq.~(\ref{eq:qme}), the state of the system $\rho(t)$, as a function of $t$, can be obtained, and the relevant quantities can subsequently be calculated. In the rest of the paper, we will employ the dimensionless system parameter $g=\frac{h}{J}$, dimensionless time $t\rightarrow \frac{Jt}{\hbar}$ and dimensionless temperature $T\rightarrow \frac{k_BT}{J}$ for describing the system and its dynamics.

\subsection{Localizable entanglement}
\label{subsec:entanglement}

Localizable entanglement (LE)~\cite{verstraete2004,popp2005,verstraete2004a,jin2004} over a selected subset of qubits in the system is defined as the maximum average entanglement localized over the selected set of qubits via local projection measurements on all the other qubits, and is expressed as 
\begin{eqnarray}
E^L_{\Omega}=\underset{\{M_{\overline{\Omega}}\}}{\max}\sum_{k}p_kE(\rho_{\Omega}^k).
\label{eq:le}
\end{eqnarray}
Here, $\{M_{\overline{\Omega}}\}$ is the complete set of single-qubit projection measurements performed over all the qubits in the set $\overline{\Omega}$, $\Omega$ is the set of selected qubits over which the localizable entanglement is to be computed such that $\Omega\cup\overline{\Omega}$ represents the entire system, $\Omega\cap\overline{\Omega}=\emptyset$, and $\rho_{\Omega}^k=\text{Tr}_{\overline{\Omega}}[\rho^k]$, with  
\begin{eqnarray}
\rho^k &=& \frac{1}{p_k}\left[\left(M_{\overline{\Omega}}\otimes I_{\Omega}\right)\rho \left(M_{\overline{\Omega}}\otimes I_{\Omega}\right)\right]
\end{eqnarray}
being the post-measured state of the system, $I_{\Omega}$ is the identity operator in the Hilbert space of the qubits in $\Omega$, and 
\begin{eqnarray}
p_k &=& \text{Tr}\left[\left(M_{\overline{\Omega}}\otimes I_{\Omega}\right)\rho \left(M_{\overline{\Omega}}\otimes I_{\Omega}\right)\right]
\end{eqnarray}
is the probability of obtaining the measurement outcome $k$ over the qubits in $\overline{\Omega}$. The definition of localizable entanglement depends on the existence of an entanglement measure $E$, referred to as the \emph{seed measure} (cf.~\cite{sadhukhan2017}), which can be computed for the post-measured state $\rho^k_\Omega$ over the subsystem $\Omega$. Depending on the situation, $E$ can either be a bipartite~\cite{verstraete2004,popp2005,verstraete2004a,jin2004,amaro2018,amaro2020a} or a multipartite entanglement measure~\cite{sadhukhan2017,streltsov2015}. Unless otherwise stated, in this paper, we focus on computing the bipartite localizable entanglement over a subset $\Omega$ of qubits forming a nontrivial loop ~\cite{kitaev2001,kitaev2003,kitaev2006,bombin2006,bombin2007} of length $|\Omega|$, and we choose negativity~\cite{peres1996,horodecki1996,zyczkowski1998,lee2000,vidal2002,plenio2005,leggio2020} as the seed measure in all our calculations (see Appendix~\ref{app:negativity} for a definition).

\begin{figure*}
    \centering
    \includegraphics[width=0.6\textwidth]{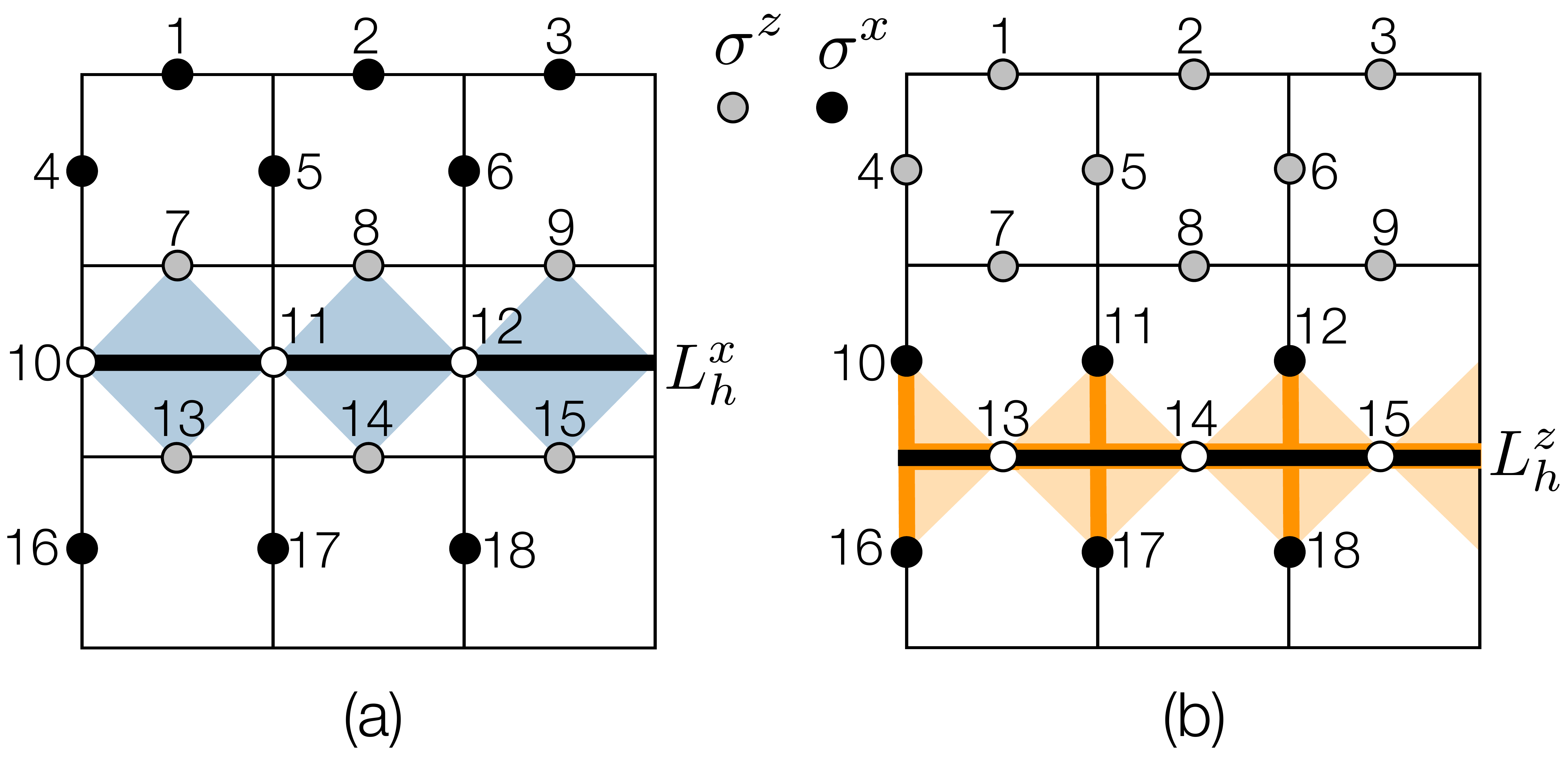}
    \caption{(Color online) \textbf{Canonical measurement setup for Kitaev code.} If the nontrivial loop represents a $L^x_{h}$ ($L^z_{h}$) operator, then the qubits that are not on $L^x_{h}$ ($L^z_{h}$), but are on the plaquette operators (vertex operators) through which $L^x_{h}$ ($L^z_{h}$) passes, are measured in $\sigma^z$ ($\sigma^x$) basis. The rest of the qubits that are not on these plaquette (vertex) operators are measured in the $\sigma^x$ ($\sigma^z$) basis. This is demonstrated for (a) $L^x_h$ (qubits $7$, $8$, $9$, $13$, $14$, $15$ are measured in $\sigma^z$ basis, and the rest in $\sigma^x$ basis) and (b) $L^z_{h}$ (qubits $10$, $11$, $12$, $16$, $17$, $18$ are measured in $\sigma^x$ basis, and the rest in $\sigma^z$ basis) in the figures. The grey (black) color of the qubits signifies  $\sigma^z$ ($\sigma^x$) measurement, while the unmeasured qubits are represented by white circles.}
    \label{fig:fig4}
\end{figure*}

It is generally difficult to compute localizable entanglement when the single-qubit measurements are to be performed over a large number $|\overline{\Omega}|$ of qubits, and analytical determination of the optimal measurement basis is possible only in few cases, eg. GHZ~\cite{sadhukhan2017,greenberger1989} and W states~\cite{sadhukhan2017,zeilinger1992,dur2000}, Dicke states~\cite{sadhukhan2017,dicke1954,kumar2017}, stabilizer states~\cite{hein2004,hein2006,amaro2018,amaro2020a}, and a number of lattice spin models with certain symmetries~\cite{verstraete2004,popp2005,verstraete2004a,jin2004,venuti2005}. In the cases of large quantum states $\rho$ where the optimization of localizable entanglement cannot be determined analytically, one can define a \emph{restricted} localizable entanglement (RLE)~\cite{amaro2018,amaro2020a,banerjee2020,banerjee2020a}, given by 
\begin{eqnarray}
E^{RL}_{\Omega}=\underset{\left\{M_{\overline{\Omega}}^\sigma\right\}}{\max}\sum_{k}p_kE(\rho_{\Omega}^k),
\label{eq:rle}
\end{eqnarray}
by confining the measurement basis for each qubit in $\overline{\Omega}$ to the eigenvectors of one of the three Pauli matrices $\sigma^x,\sigma^y$, and $\sigma^z$, where $\{M_{\overline{\Omega}}^\sigma\}$ is the complete set of all possible Pauli measurement configurations on all the qubits in $\overline{\Omega}$. The definitions of LE and RLE suggests that
\begin{eqnarray} 
E^L_\Omega\geq E^{RL}_{\Omega},
\label{eq:hierarchy1}
\end{eqnarray}
where the same seed measure is chosen for computing both the LE and the RLE. While the equality in Eq.~(\ref{eq:hierarchy1}) occurs only in a few cases~\cite{verstraete2004,popp2005,verstraete2004a,jin2004,sadhukhan2017,amaro2018,amaro2020a,banerjee2020,banerjee2020a}, there exists quantum states where  $\left|E^L_\Omega-E^{RL}_{\Omega}\right|$ is so small that the LE can be well-approximated using the RLE~\cite{banerjee2020,banerjee2020a}, and analytical expressions for the RLE can also be determined. 

\begin{figure}
    \centering
    \includegraphics[width=0.8\linewidth]{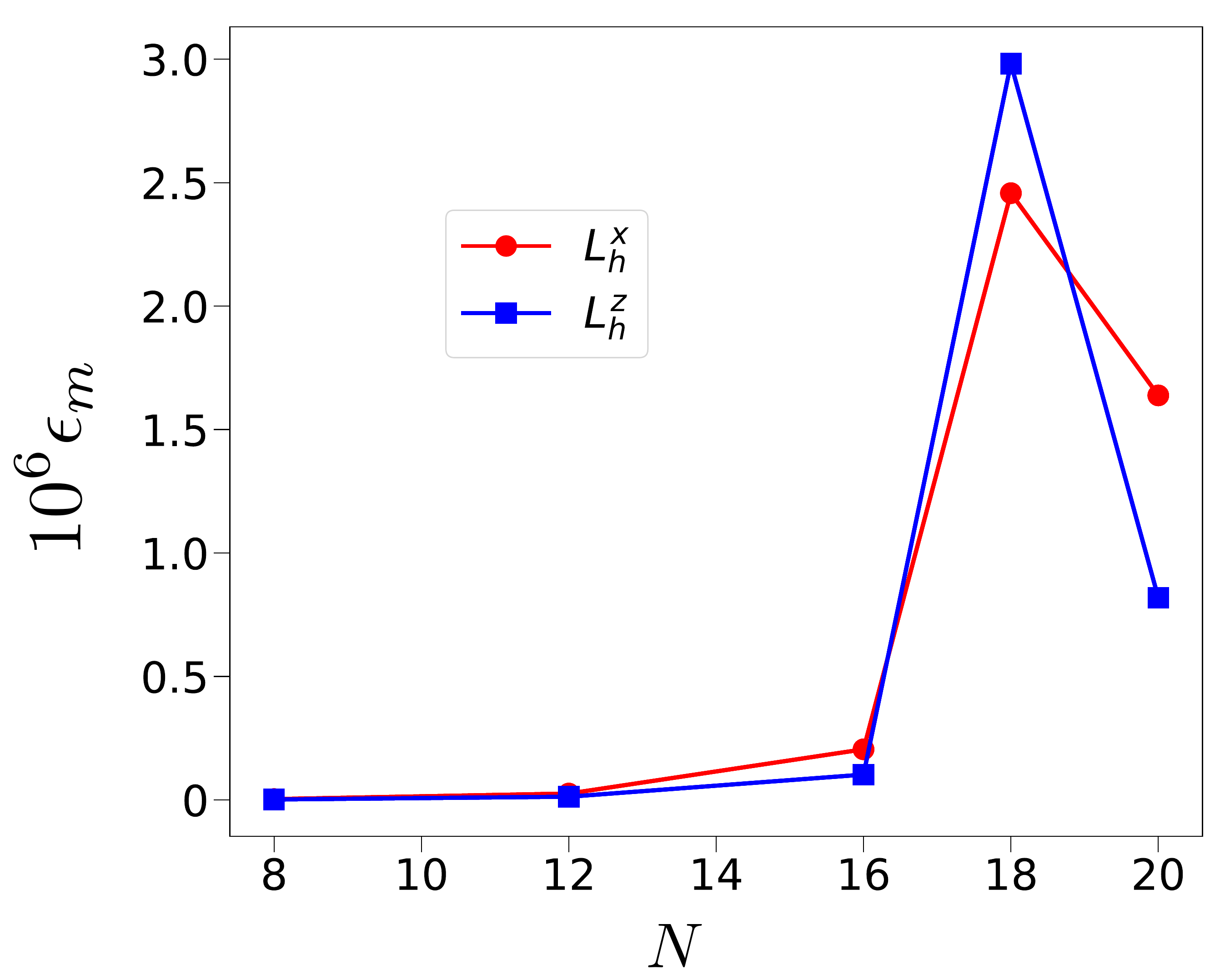}
    \caption{(Color online) \textbf{Estimated maximum error for $E^{\prime\prime}_\Omega$.} Variation of $\epsilon_m$ as a function of $N$ for Kitaev code, where localizable entanglement is computed over $\Omega\equiv L^x_h$ and $L^z_h$. The maximum error occurs in the case of  $9$-plaquette toric code, although all values of $\epsilon_m\sim 10^{-6}$ or lower. Both the horizontal and the vertical axes are dimensionless.}
    \label{fig:fig5}
\end{figure}

Note that the optimization in RLE requires consideration of $3^{|\overline{\Omega}|}$ possible configurations of Pauli measurement setups, which can be difficult in situations where $|\overline{\Omega}|$ is a large number. However, one can choose a specific Pauli measurement setup from the complete set of $3^{|\overline{\Omega}|}$ configurations, where the choice depends on the structure of the quantum state (cf.~\cite{hein2004,hein2006,amaro2018,amaro2020a}), or the symmetry of the system (cf.~\cite{venuti2005}). While such a choice does not guarantee the optimized value of the RLE for all possible values of the varying parameters in the system, a judicious choice would provide a lower bound, $E^{\prime}_{\Omega}$, of the RLE, such that the hierarchy in Eq.~(\ref{eq:hierarchy1}) becomes 
\begin{eqnarray} 
E^L_\Omega\geq E^{RL}_{\Omega}\geq E^\prime_\Omega,
\label{eq:hierarchy2}
\end{eqnarray}
where the bound 
\begin{eqnarray}
E^\prime_\Omega=\sum_{k=0}^{2^{|\overline{\Omega}|}-1}p_kE(\rho^k_\Omega)
\label{eq:pauli_bound}
\end{eqnarray}
can be analytically computed in some cases (cf.~\cite{amaro2018,amaro2020a}). Although the existence of a Pauli measurement setup over the qubits in $\overline{\Omega}$ corresponding to small value of $\left|E^{RL}_{\Omega}-E^\prime_\Omega\right|$ is not guaranteed, in the occasions where it exists, there are situations where the variations of $E^\prime_\Omega$ as functions of the relevant parameters are qualitatively same as the variations of the RLE with the same parameters~\cite{amaro2018,amaro2020a,banerjee2020,banerjee2020a}. We shall see specific advantages of this in investigating the topological to nontopological QPT in Section~\ref{subsec:canonical}. 

\begin{figure*}
    \centering
    \includegraphics[width=0.9\textwidth]{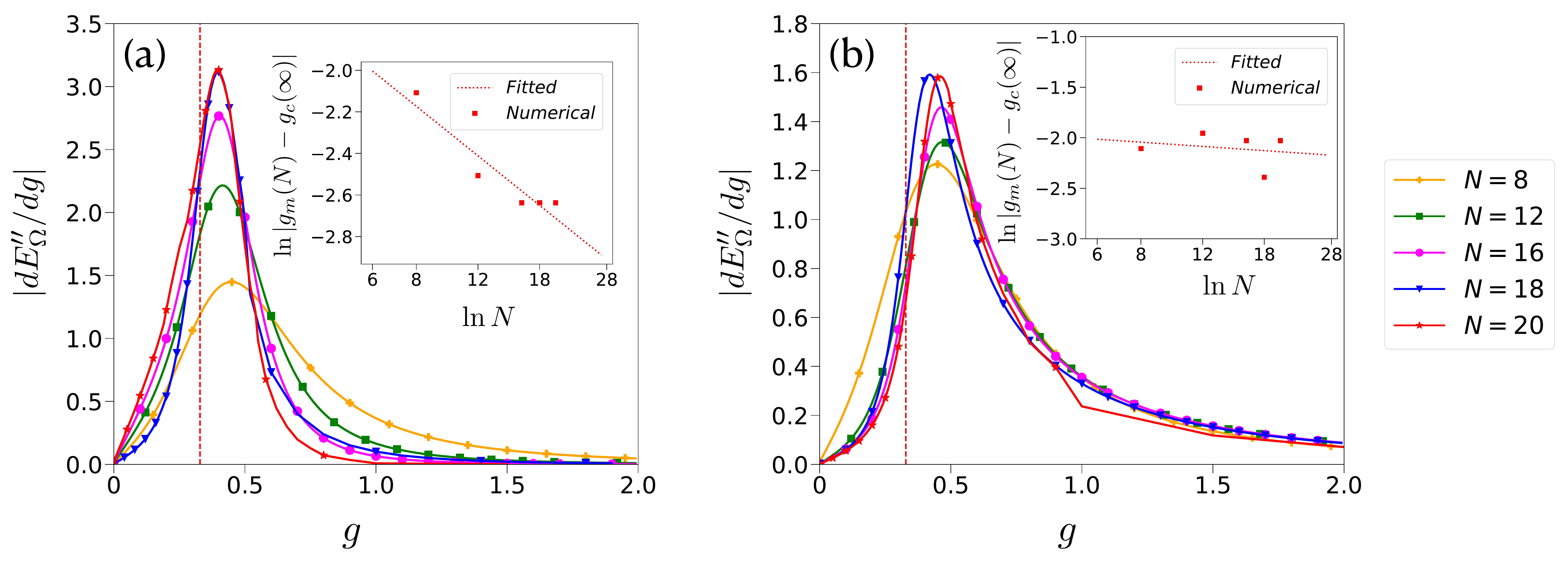}
    \caption{(Color online) \textbf{QPT using lower bound of localizable entanglement in Kitaev code.} Variations of $\left|\frac{dE^{\prime\prime}_\Omega}{dg}\right|$ as a function of $g$ across the QPT point of a Kitaev code under parallel magnetic field for different system sizes $N$, when $\Omega$ is chosen to be a nontrivial loop representing (a) $L^x_h$ and (b) $L^z_h$. The Kitaev code is considered over a lattice of $N_P^h\times N_P^v$ architecture (see Section~\ref{subsubsec:kitaev}) with $N^{h,v}_P$ being the number of plaquettes in the horizontal and vertical directions, where the system size $N$ takes the values $8$  $(2\times 2)$, $12$ $(2\times 3)$, $16$ $(2\times 4)$, $18$ $(3\times 3)$, and $20$ $(2\times 5)$. (Insets of (a) and (b)) The variation of $\ln\left|g_m(N)-g_c(\infty)\right|$ as a function of $\ln N$. The dotted line represents the fitted curve according to Eq.~(\ref{eq:le_scaling}), with (a) $\alpha=0.38(6)$, $\nu=0.58(8)$ for $\Omega\equiv L^x_h$ and (b) $\alpha=0.10(2)$, $\nu=0.16(0)$ for $\Omega\equiv L^z_h$. Both the horizontal and the vertical axes in all figures are dimensionless.}
    \label{fig:fig6}
\end{figure*}

The computation of LE and its lower bound is feasible in experiments via using appropriate entanglement witness operators~\cite{horodecki2009,guhne2009,terhal2002,guhne2002,guhne2005,alba2010,amaro2020} for the post-measured states on the subsystem $\Omega$. An entanglement witness operator $W$ indicate the entanglement status of a quantum state $\rho$ via its expectation value $w=\text{Tr}[\rho W]$ in the state. If $w<0$, the state $\rho$ is an entangled state. Given a specific entanglement measure $E$, a lower bound of the entanglement content in the state $\rho$ can also be obtained as a solution of the optimization problem~\cite{eisert2007,guhne2007,guhne2008}
\begin{eqnarray}
E_{\min}(w) &=& \inf E(\rho) 
\label{eq:witness_bound}
\end{eqnarray}
subject to $w=\text{Tr}[\rho W]$, $\rho>0$, and $\text{Tr}[\rho]=1$. Assuming that a witness operator $W$ exists such that its expectation value $w_k=\text{Tr}[\rho_\Omega^k W]$ in the post-measures state $\rho^k_\Omega$ corresponding  to the measurement outcome $k$ provides a lower bound $E_{\min}(w_k)$ of $E(\rho^k_\Omega)$, Eq.~(\ref{eq:pauli_bound}) leads to 
\begin{eqnarray}
E^\prime_\Omega\geq \sum_{k=0}^{2^{|\Omega|}-1}p_k E_{\min}(w_k)=E^w_\Omega,
\label{eq:witness_lb}
\end{eqnarray}
such that\begin{eqnarray} 
E^L_\Omega\geq E^{RL}_{\Omega}\geq E^\prime_\Omega\geq E^w_\Omega.
\label{eq:hierarchy3}
\end{eqnarray}
We point out here that identifying appropriate entanglement witness for states of a quantum system with varying system parameters is a challenging problem, as the state of the system may change with the value of the system parameter. Moreover, the calculation of $E^w_\Omega$ still requires a measurement on the qubits in $\overline{\Omega}$ according to a chosen Pauli measurement setup, and in turn, determination the expectation value of the chosen $W$ in a total of $2^{|\overline{\Omega}|}$ post-measured states of the subsystem $\Omega$. Possible solutions to these challenges have been proposed in~\cite{amaro2018,amaro2020,amaro2020a} in the specific case of  topological quantum codes. We discuss this in detail in  Section~\ref{subsec:qpt_witness}. 

Unless otherwise mentioned, we focus on $E^\prime_\Omega$ and $E^{w}_\Omega$ for the topological quantum codes considered in this paper. More specifically, for larger systems, we compute an approximation of $E^\prime_\Omega$, which we introduce in Section~\ref{subsubsec:rolb}.

\begin{figure*}
    \centering
    \includegraphics[width=0.6\textwidth]{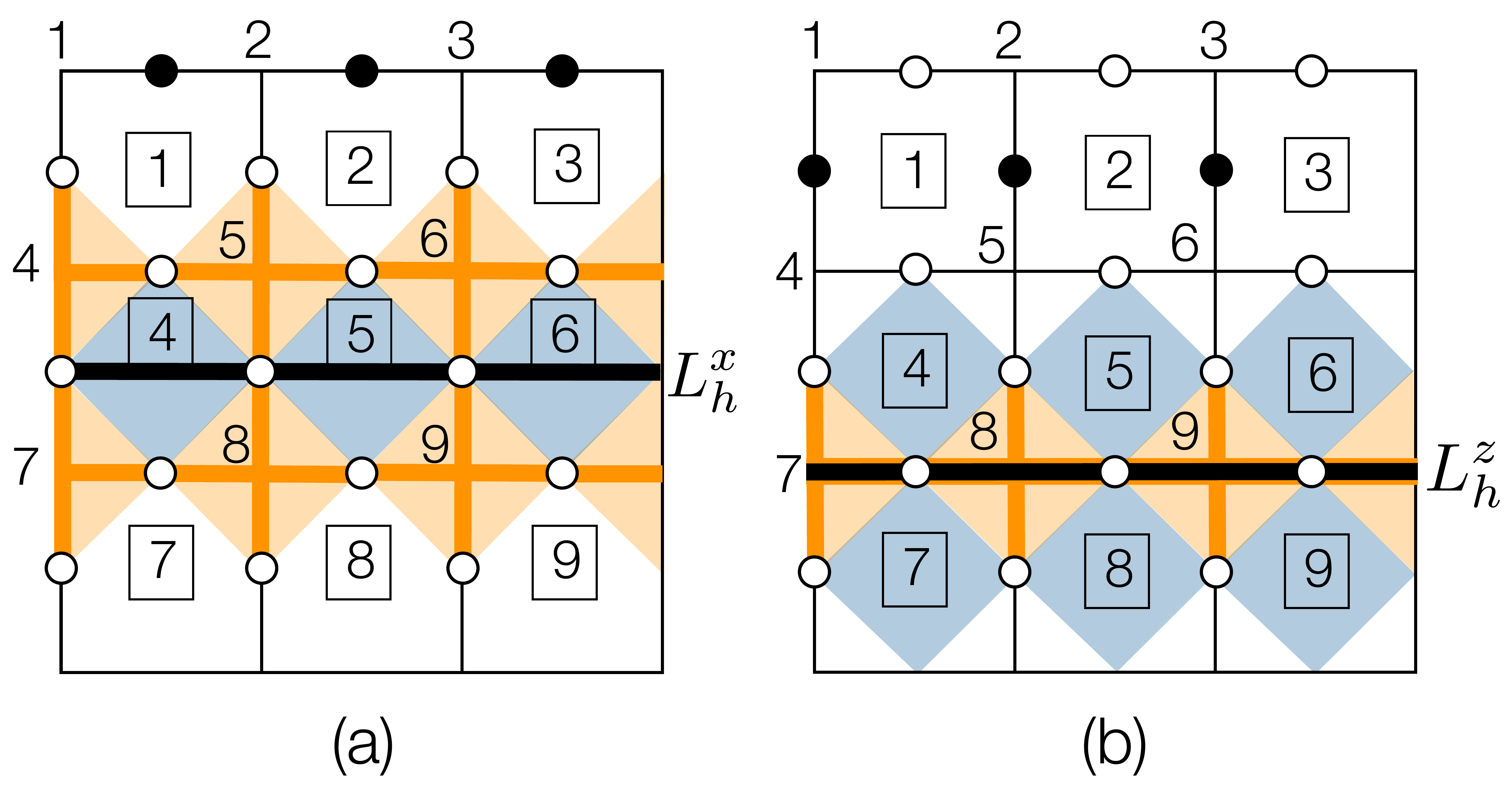}
    \caption{(Color online) \textbf{Constructing the local witness operators in Kitaev code.} The plaquette and the vertex stabilizer operators, a subset of which  can potentially contribute in building a local witness operators corresponding to a nontrivial loop representing (a) $L^x_h$ and (b) $L^z_h$, are shown for a $9$-plaquette Kitaev code. The qubits denoted by the white circles are the ones on which the witness operators may have support. The labels of the plaquettes are denoted by boxed numbers, while numbers alone denote the vertices. See Eqs.~(\ref{eq:kitaev_stabilizers}) for the descriptions of the plaquette and the vertex stabilizer operators.}
    \label{fig:fig7}
\end{figure*}

\section{On the Optimal Basis for Localizable Entanglement}
\label{sec:optimal_basis}

We now investigate the behavior of LE over a nontrivial loop in a topological quantum code when the strength of the parallel magnetic field is varied across a topological to nontopological QPT point, and discuss the corresponding finite-size scaling for LE. Here and in the rest of the paper, unless otherwise stated, we compute the entanglement, as quantified by the negativity, over the post-measured states on $\Omega$ in the $1:\text{rest}$ partition, i.e.,  a bipartition of a single qubit, and the rest of the qubits in $\Omega$. The symmetry of the system ensured that the entanglement is invariant with respect to the choice of the single qubit in $\Omega$. Also, for the purpose of demonstration, we always localize entanglement over a subsystem $\Omega$ of qubits forming a nontrivial loop corresponding to the logical operator $L^{\alpha}_h$, $\alpha=x,z$. 

\begin{table*}[]
\centering 
Construction of the witness operators \\ 
\begin{tabular}{p{0.075\linewidth} p{0.15\linewidth} p{0.15\linewidth} p{0.15\linewidth} p{0.15\linewidth} p{0.15\linewidth}}
\hline 
     & $N=8$ & $N=12$ &  $N=16$ & $N=18$ & $N=20$ \\
\hline      
    & & & & & \\
    $L^x_h$ & $\vcenter{\hbox{\includegraphics[width=5em]{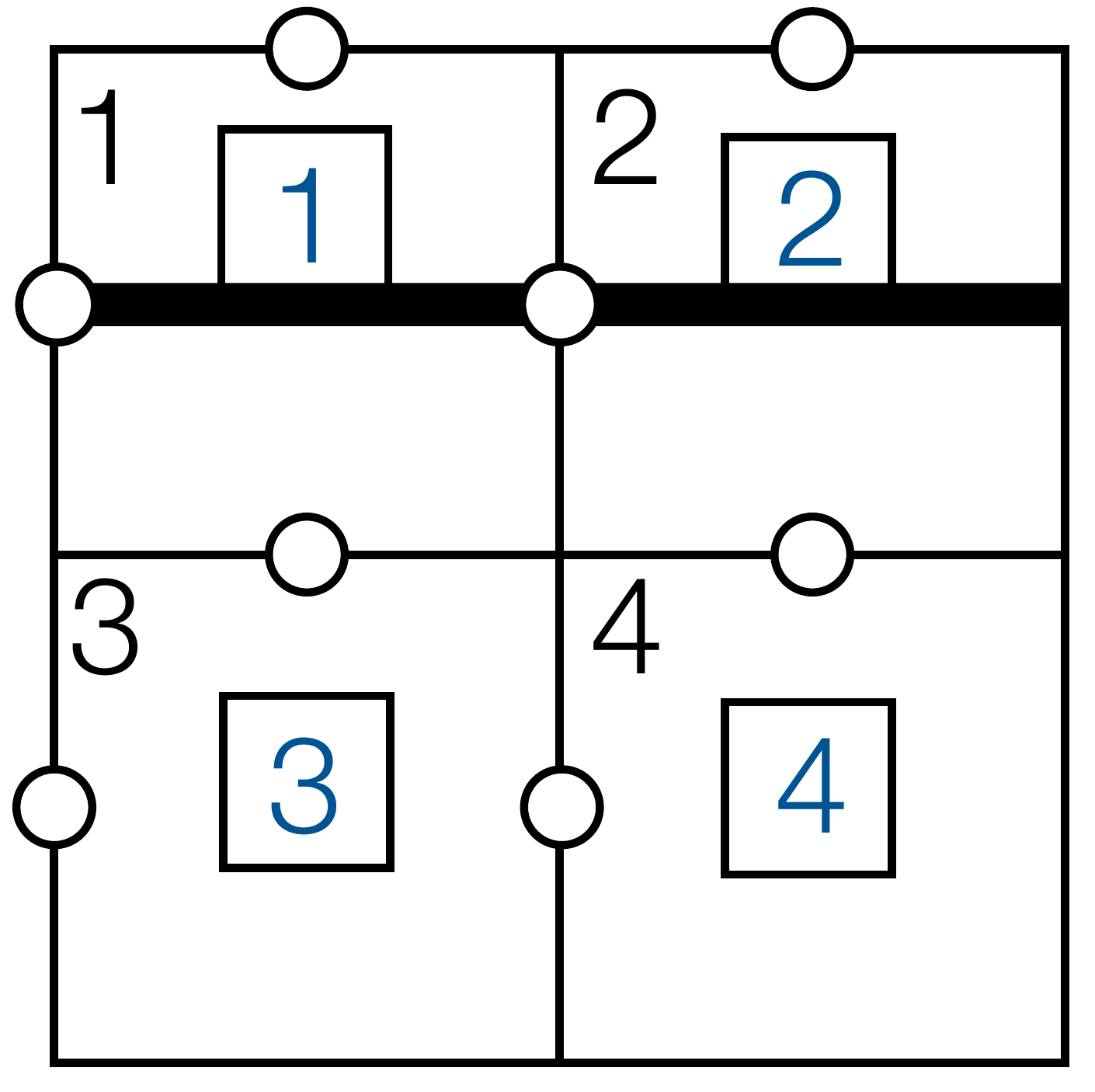}}}$ & $\vcenter{\hbox{\includegraphics[width=6em]{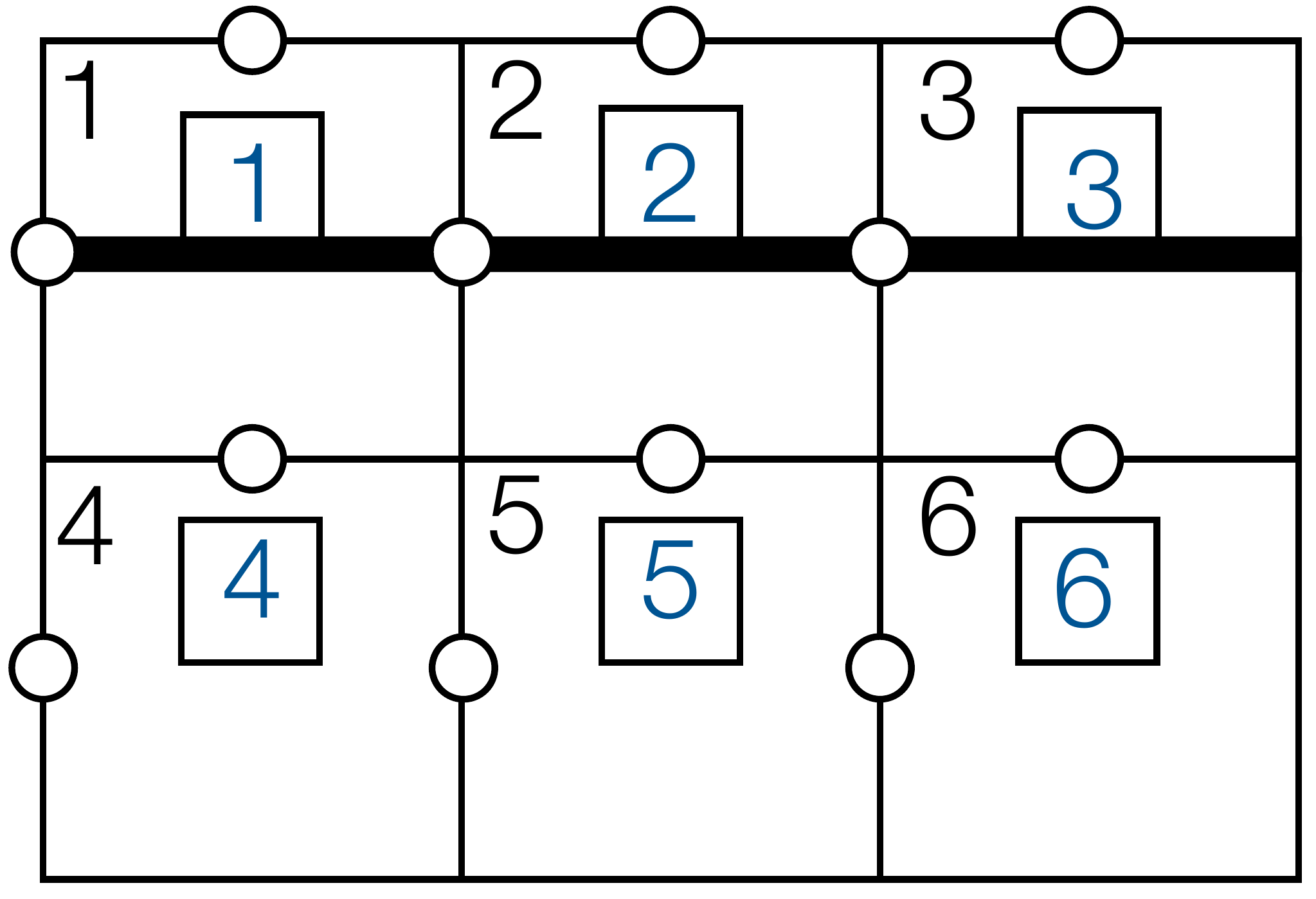}}}$ &  $\vcenter{\hbox{\includegraphics[width=7em]{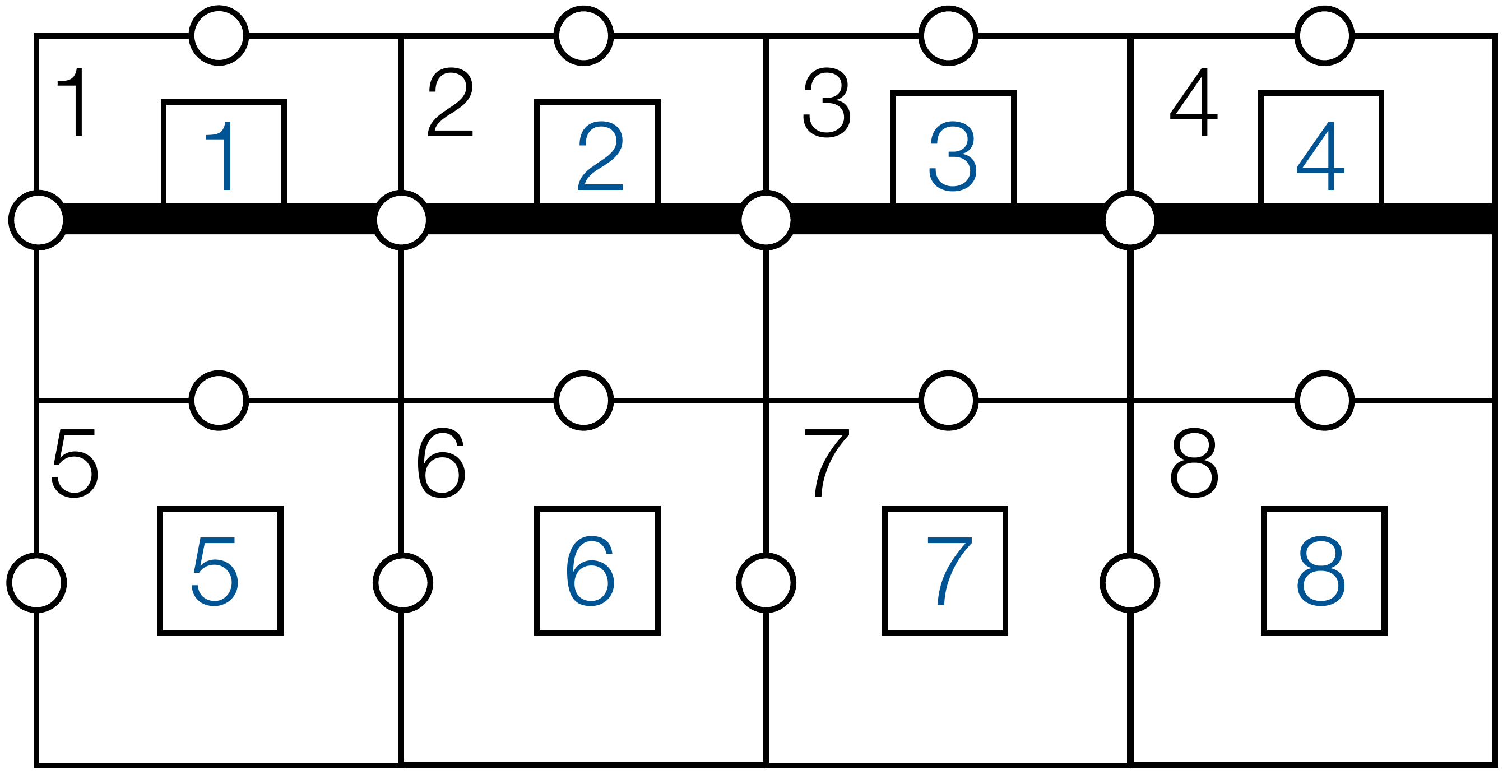}}}$ & $\vcenter{\hbox{\includegraphics[width=6em]{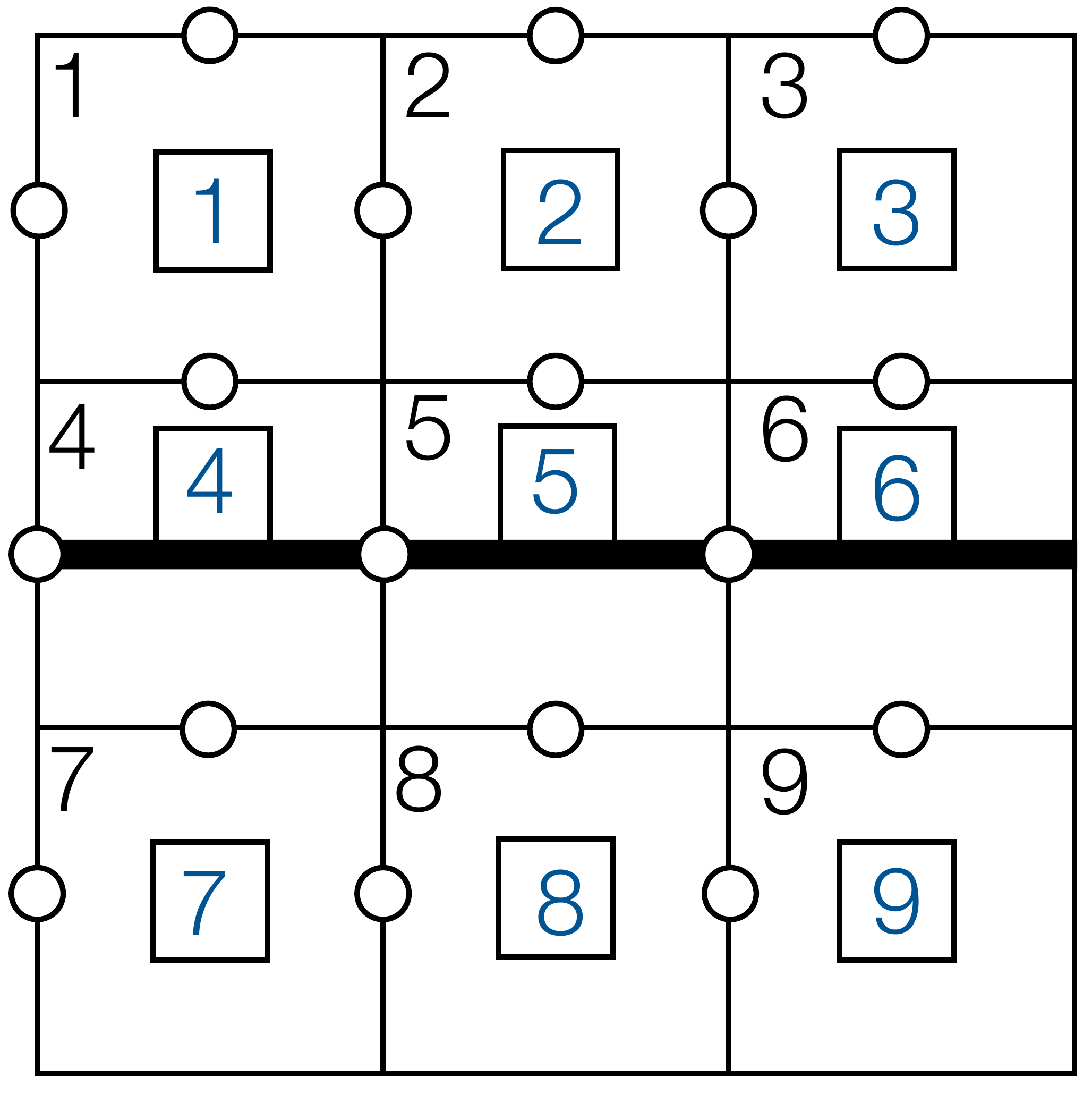}}}$ & $\vcenter{\hbox{\includegraphics[width=8em]{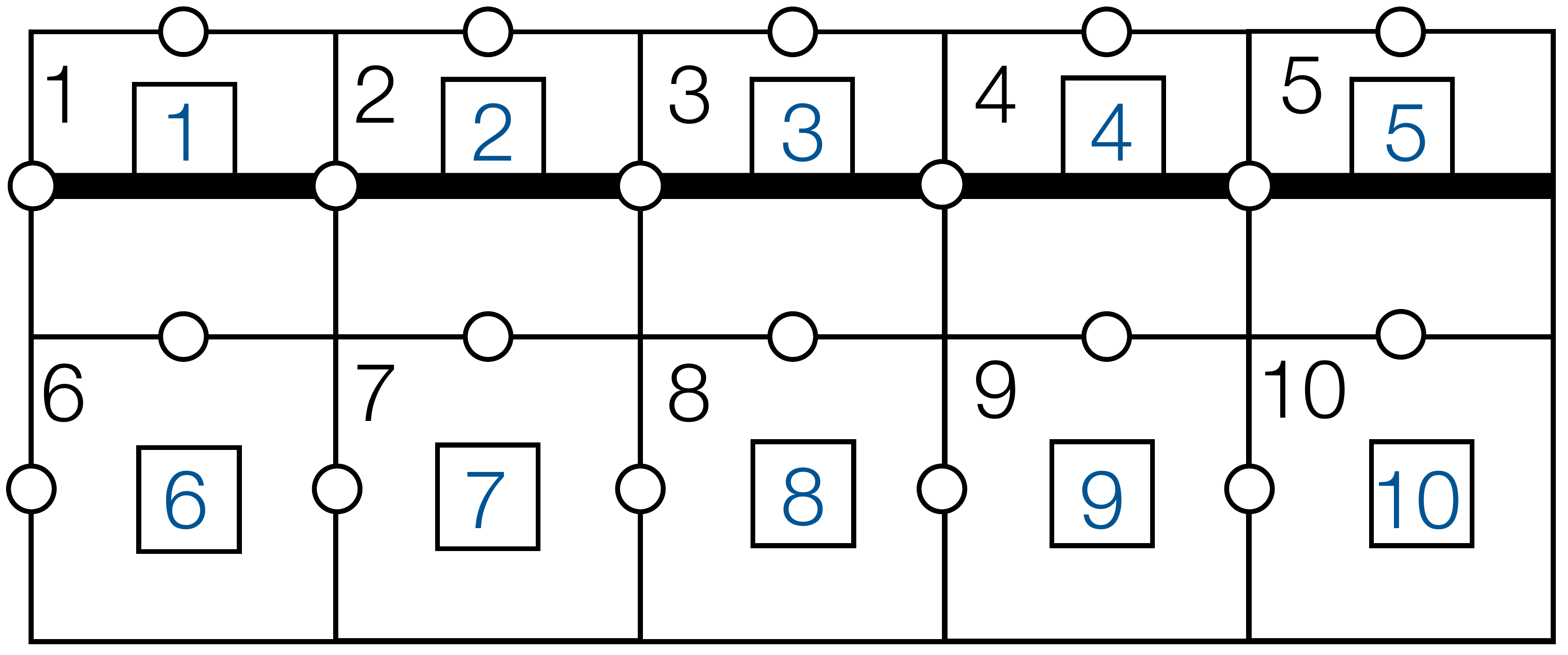}}}$ \\ 
    $\{s_j\}$ & $s_1=V_3V_4$,\newline $s_2=P_2$  & $s_1=V_4V_5V_6$,\newline $s_2=P_2P_3$, \newline $s_3=P_3$ & $s_1=V_5V_6V_7V_8$,\newline $s_2=P_2P_3P_4$, \newline $s_3=P_3P_4$, \newline $s_4=P_4$ & $s_1=V_7V_8V_9$,\newline $s_2=P_5P_6$, \newline $s_3=P_6$ & $s_1=V_6V_7V_8V_9V_{10}$,\newline $s_2=P_2P_3P_4P_5$, \newline $s_3=P_3P_4P_5$, \newline $s_4=P_4P_5$, \newline $s_5=P_5$ \\
\hline      
    & & & & & \\
    $L^z_h$ & $\vcenter{\hbox{\includegraphics[width=5em]{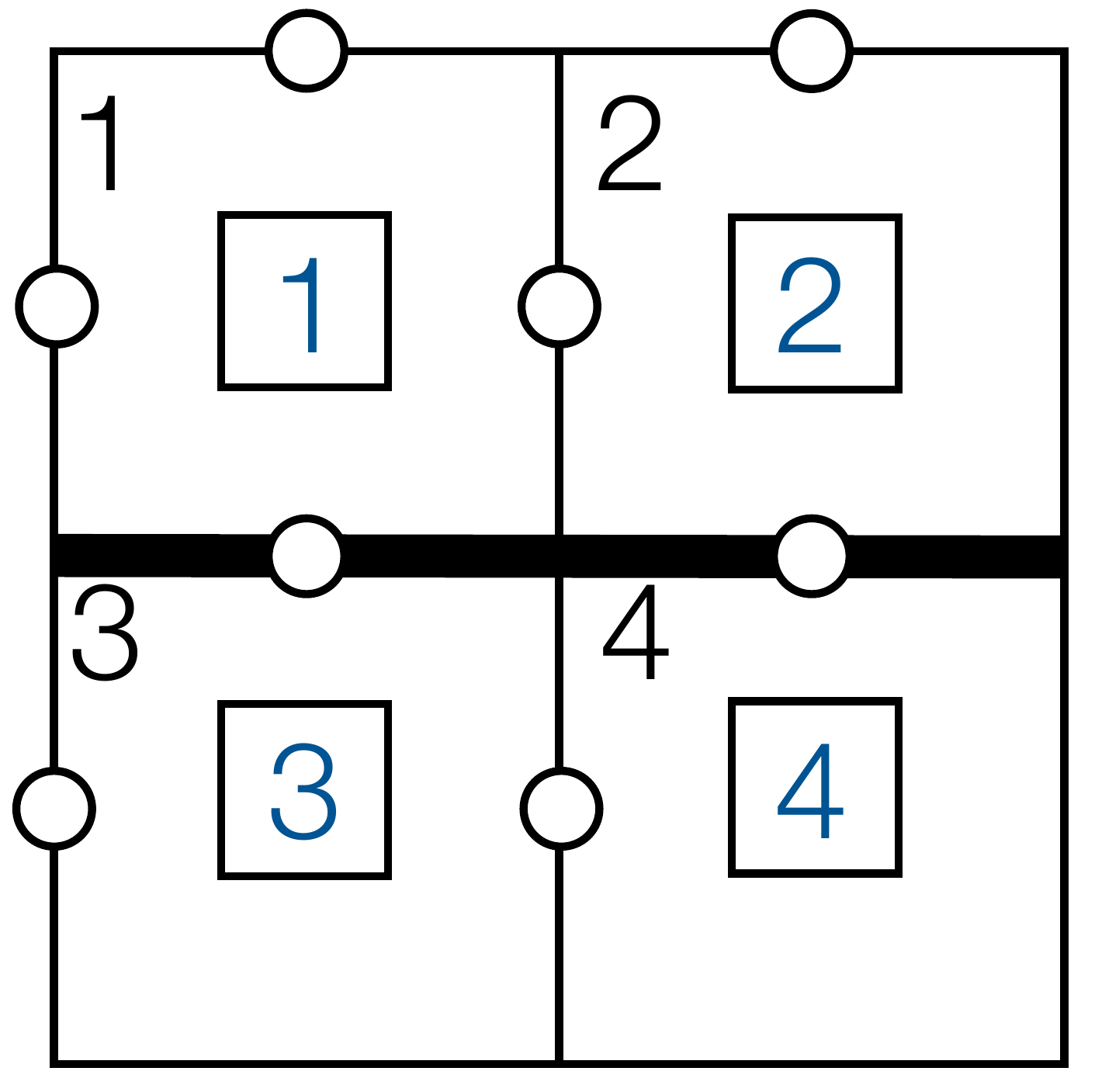}}}$ & $\vcenter{\hbox{\includegraphics[width=6em]{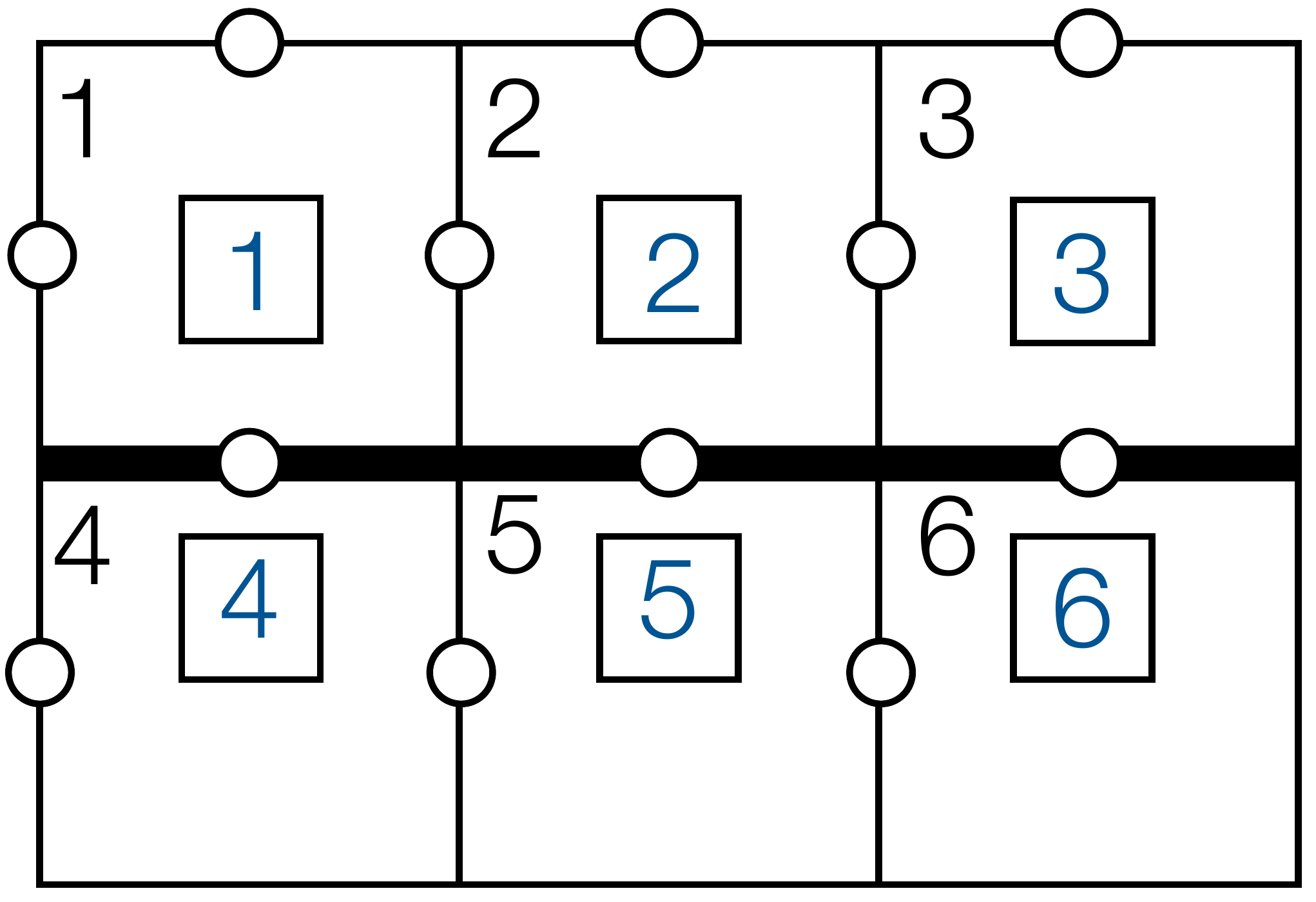}}}$ &  $\vcenter{\hbox{\includegraphics[width=7em]{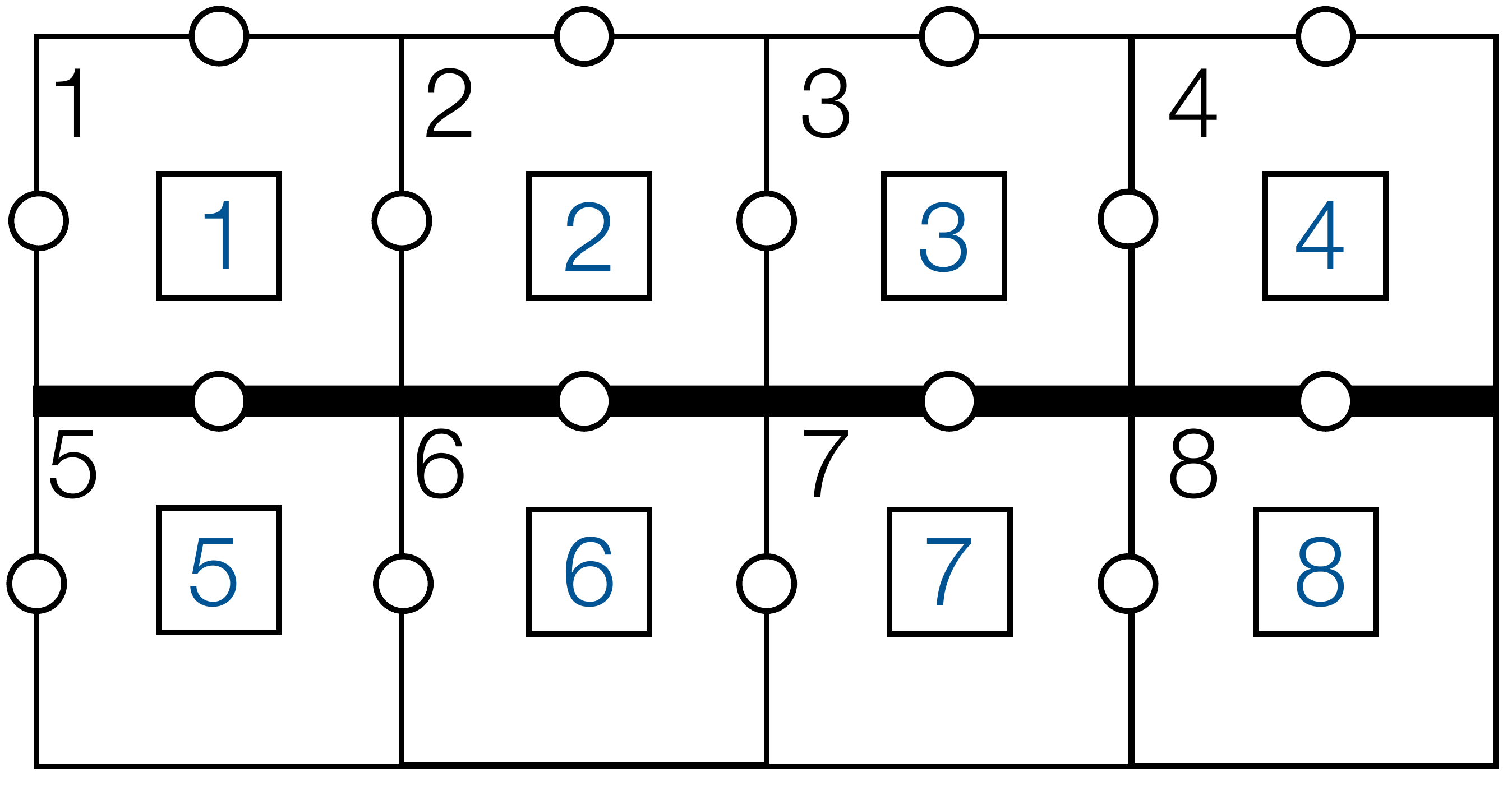}}}$ & $\vcenter{\hbox{\includegraphics[width=6em]{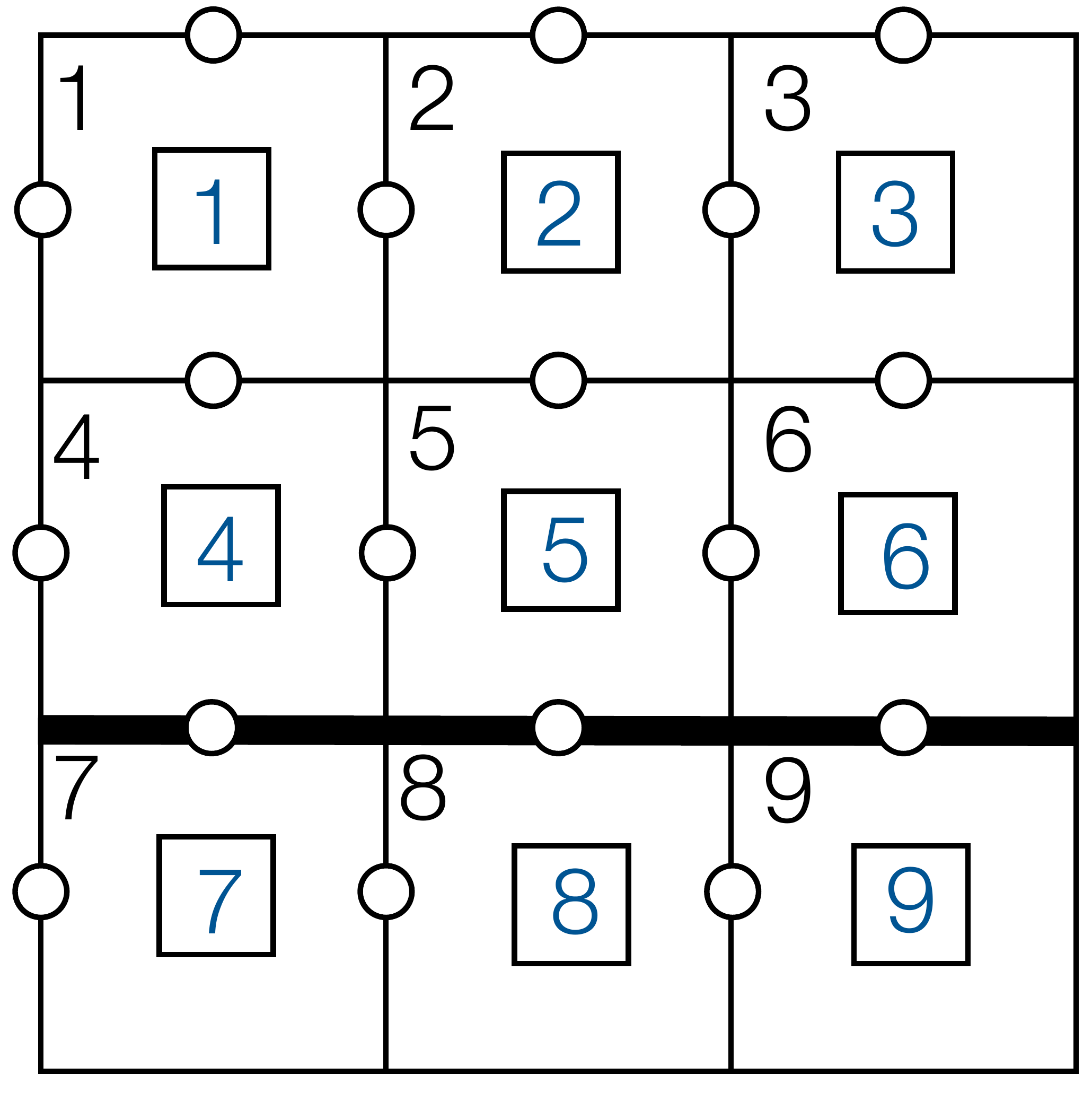}}}$ & $\vcenter{\hbox{\includegraphics[width=8em]{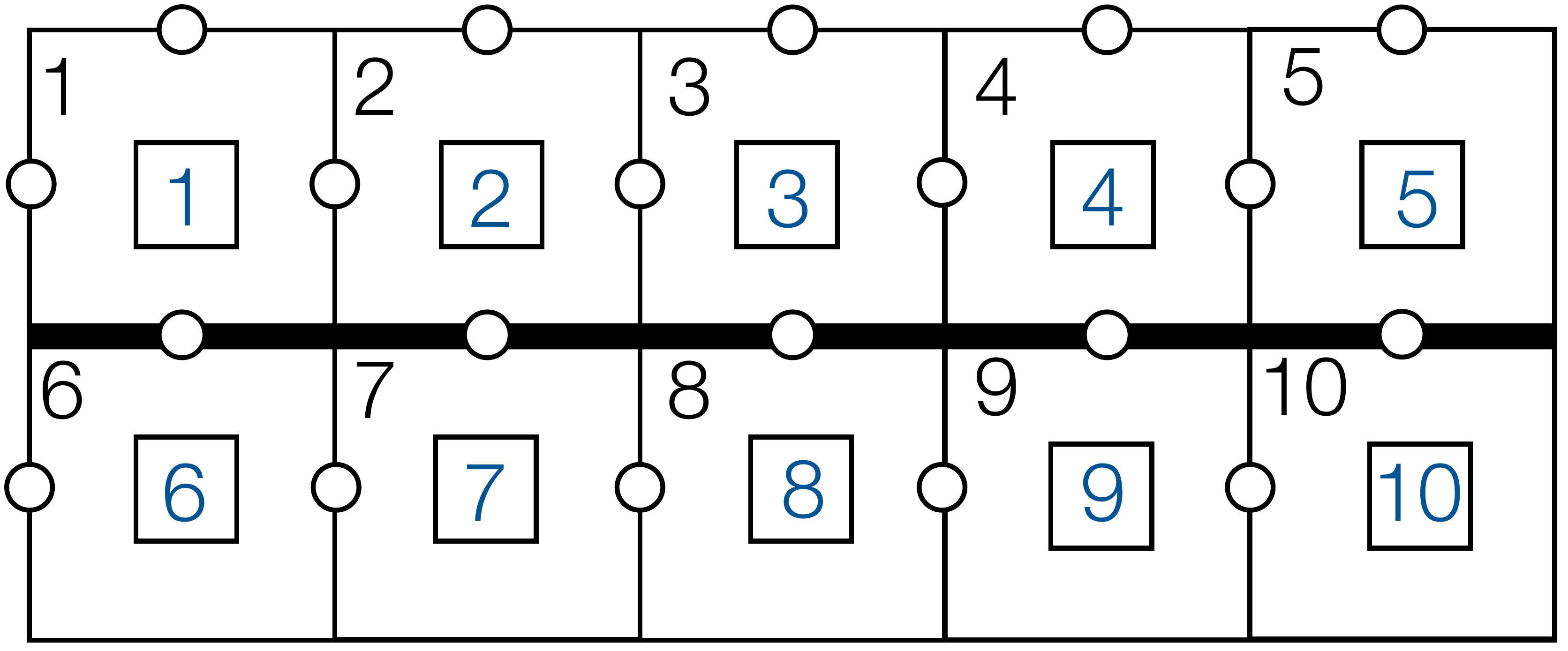}}}$ \\   
    $\{s_j\}$ & $s_1=P_1P_2$,\newline $s_2=V_4$  & $s_1=P_1P_2P_3$,\newline $s_2=V_5V_6$, \newline $s_3=V_6$ & $s_1=P_1P_2P_3P_4$,\newline $s_2=V_6V_7V_8$, \newline $s_3=V_7V_8$, \newline $s_4=V_8$ & $s_1=P_4P_5P_6$,\newline $s_2=V_8V_9$, \newline $s_3=V_9$ & $s_1=P_1P_2P_3P_4P_5$,\newline $s_2=V_7V_8V_9V_{10}$, \newline $s_3=V_8V_9V_{10}$, \newline $s_4=V_9V_{10}$, \newline $s_5=V_{10}$ \\    
\end{tabular}
\caption{List of stabilizer operators used to construct $W_\Omega$ (Eq.~(\ref{eq:local_witness})) in the case of the Kitaev code on rectangular and square lattices, when $\Omega$ is chosen to be a nontrivial loop (represented by thick black lines) representing $L^x_h$ and $L^z_h$.  Periodic boundary conditions are assumed along both the horizontal and the vertical directions of all the lattices. The plaquettes are labeled with boxed numbers, while numbers alone denote the vertices. For ease of notations, we have represented the vertex operators $S_v$ with $V_v$, and the plaquette operators $S_p$ with $P_p$, $p$ and $v$ respectively representing the plaquette and the vertex indices. See Eq.~(\ref{eq:kitaev_stabilizers}) for the definitions of $S_p$ and $S_v$. See also Figure~\ref{fig:fig7}.}
\label{tab:tab1}
\end{table*}

Let us first consider the Kitaev model in a parallel field (Eq.~(\ref{eq:kitaev_hamiltonian})) on a square lattice with four plaquettes in $2\times 2$ architecture, where the nontrivial loop in the $g\rightarrow 0$ limit of the Hamiltonian (\ref{eq:kitaev_hamiltonian}) is constituted of two qubits constructing the region $\Omega$ (see also Table~\ref{tab:tab1}). 
We diagonalize $H_K$ to obtain the ground state of the system, and numerically determine $E^L_{\Omega}$ and  $E^{RL}_{\Omega}$ as functions of the magnetic field strength $g$. Our data suggests that for all values of $g$ over a considerably wide range $[0,2.0]$ across the QPT point, the optimization of the LE takes place in the Pauli basis, which is indicated by the vanishing $\left|E^{L}_\Omega-E^{RL}_\Omega\right|$ (see Section~\ref{subsec:entanglement}) for all $g$. This is demonstrated by the coincidence of the graphs for $E^L_{\Omega}$ and $E^{RL}_{\Omega}$ in Figure~\ref{fig:fig3}(a) and (b), for both types of the nontrivial loops, $L^x_{h}$ and $L^z_{h}$. However, the computation of RLE quickly becomes intractable due to the exponential increase in the number of possible Pauli measurement setup, thereby making it difficult to probe the topological to nontopological QPT via LE and RLE.


\subsection{Canonical measurement setup}
\label{subsec:canonical}

In view of the above discussion, we now investigate the performance of $E^\prime_{\Omega}$ as a lower bound of RLE (see Section~\ref{subsec:entanglement}), for which a specific Pauli measurement setup needs to be chosen. To make this choice, we note that the ground state of $H_K$ is equivalent to a graph state~\cite{hein2006} via a set of local Clifford unitary operations~\cite{lang2012,amaro2018,amaro2020a,HK_inprep} (see Appendix~\ref{app:canonical_setup} for a detailed discussion). With an appropriate choice of the set of qubits on which these Clifford unitaries are applied, it can be ensured that the local unitary equivalent graph state has a connected \emph{star graph} over the chosen region $\Omega$ representing a nontrivial loop on the Kitaev code~\cite{lang2012,HK_inprep}. Since the bipartite entanglement in a graph state over all $1:\text{rest}$ bipartitions is maximum as long as the graph state corresponds to a connected graph~\cite{hein2004,verstraete2003,hein2006,HK_inprep}, a Pauli measurement setup that leaves the connected star graph unchanged over $\Omega$ after measurement can lead to $E^{\prime}_\Omega=E^{RL}_\Omega$. There may exist more than one such Pauli measurement setups, and any such Pauli measurement setup for the graph state, via the Clifford unitary transformations that connect the graph state with the ground state of $H_K$, can provide a potential \emph{canonical} measurement setup for the topological code (see Appendix~\ref{app:canonical_setup} for a detailed discussion with examples). For the purpose of demonstration in this paper, we choose the setup where the measurement basis corresponding to it on different qubits in $\overline{\Omega}$ can be characterized only by the relative positions of the qubits with respect to the nontrivial loop operator of the type $L^\alpha_{h}$, $\alpha=x,z$, and the plaquette or the vertex operators through which it passes, using a simple set of rules. These rules are the following (see Figure~\ref{fig:fig4}). 
\begin{enumerate}
\item[(a)] To determine the localizable entanglement over a nontrivial loop representing the logical operator $L^x_{h}$ ($L^z_{h}$), a qubit which is not on $L^x_{h}$ ($L^{z}_{h}$) but is situated on the plaquette operators (vertex operators) through which $L^x_{h}$ ($L^z_{h}$) passes, is measured in the $\sigma^z$ ($\sigma^x$) basis. 
\item[(b)] All other qubits are measured in the $\sigma^x$ ($\sigma^z$) basis. 
\end{enumerate} 
Note that rules similar to $L^{x}_h$ $(L^z_h)$ apply for the measurement setup corresponding to the nontrivial loop $\Omega$ representing $L^{x}_v$ $(L^z_v)$. See Appendix~\ref{app:canonical_setup} for details. It is also important to mention here that the fact that the canonical measurement setup keeps the connected star graph over the region $\Omega$ unchanged after measurement is crucial for using the same measurement setup to develop and compute a witness-based lower bound. See Section~\ref{subsec:qpt_witness} and Appendix~\ref{app:witness} for a discussion.

It is now logical to ask how this canonical measurement setup  performs once the system is perturbed with the parallel magnetic field (i.e., $g>0$), which can be investigated by computing $\left|E^{RL}_\Omega-E^\prime_\Omega\right|$ as a function of $g$. We observe that a dichotomy between the behaviors of $E_\Omega^\prime$ over  $\Omega\equiv L^x_{h}$ and $\Omega\equiv L^z_{h}$ exists in the case of the four-plaquette toric code. While $\left|E^{RL}_\Omega-E^\prime_{\Omega}\right|=0$ corresponding to  $\Omega\equiv L^x_{h}$ for all values of $g\in [0,2]$, in the case of $\Omega\equiv L^{z}_{h}$, $\left|E_\Omega^{RL}-E^\prime_\Omega\right|=0$ only at $g=0$. With increasing $g$,  $E^\prime_\Omega < E^{RL}_\Omega$. This implies that the canonical measurement setup is sub-optimal in the case of $g>0$ for $L^z_{h}$, while it remains optimal for $L^x_{h}$ even when $g$ increases. However, it is clear from the qualitative behavior of $E^\prime_{\Omega}$ depicted in Figure~\ref{fig:fig3}(a) and (b) that for both the cases of $\Omega\equiv L^x_{h}$ and $\Omega\equiv L^z_{h}$, $E^\prime_{\Omega}$ reliably mimics the behavior of the RLE as a function of $g$ across the QPT point $g_c$ at least for the case of $N_P=4$. With relation to the inequivalence between the entanglement localized over a nontrivial loops of $L^x_h$ and $L^z_h$ type, we also point out that the applied magnetic field is taken to be in the $z$-direction only.

Note that the computation of $E^\prime_\Omega$ still involves computation of an entanglement measure over $2^{|\Omega|}$-dimensional density matrices corresponding to each of  $2^{|\overline{\Omega}|}$ measurement outcomes. Since the number of qubits in the system increases rapidly with increasing the lattice size, the numbers $2^{|\overline{\Omega}|}$ and $2^{|{\Omega}|}$ also grow fast with the lattice.  Therefore the determination of $E^\prime_{\Omega}$ becomes computationally demanding with increasing system size. Since we aim to investigate the QPT via LE, we propose the following approximation to reduce the computational resource required to calculate $E^\prime_\Omega$, so that it can be computed for higher system sizes. 


\subsubsection{Approximating the lower bound for larger systems}
\label{subsubsec:rolb}

To reduce the computational complexity of $E^{\prime}_\Omega$, we note from our numerical analysis that among the full set of $2^{|\overline{\Omega}|}$ measurement outcomes, not all have considerable probability of occurrence. Depending on this observation, we approximate $E^\prime_\Omega$ as 
\begin{eqnarray}
E^{\prime\prime}_{\Omega}=\sum_{k\in\mathcal{K}}p_k E(\rho^k_{\Omega})\;\;\text{ s.t. } p_k>p_c\;\;\forall k\in\mathcal{K}. 
\label{eq:proposed_lower_bound}
\end{eqnarray}
where only a \emph{preferred set} $\mathcal{K}$ of  measurement outcomes occurring with a probability greater than a \emph{threshold} value $p_c$ are considered. Note here that $E^{\prime\prime}_\Omega$ is, by definition, a lower bound of $E^\prime_\Omega$, i.e., $E^{\prime\prime}_\Omega\leq E^\prime_\Omega$. Note also that the value of $E^{\prime\prime}_\Omega$ is specific to the canonical measurement setup $E^\prime_{\Omega}$. The motivation behind such an approximation may be justified looking into the effect of the Pauli measurements on a graph state, and the connection between the stabilizer states with graph states~\cite{HK_inprep,elliott2008,elliott2009}. 

\begin{figure*}
    \centering
    \includegraphics[width=0.9\textwidth]{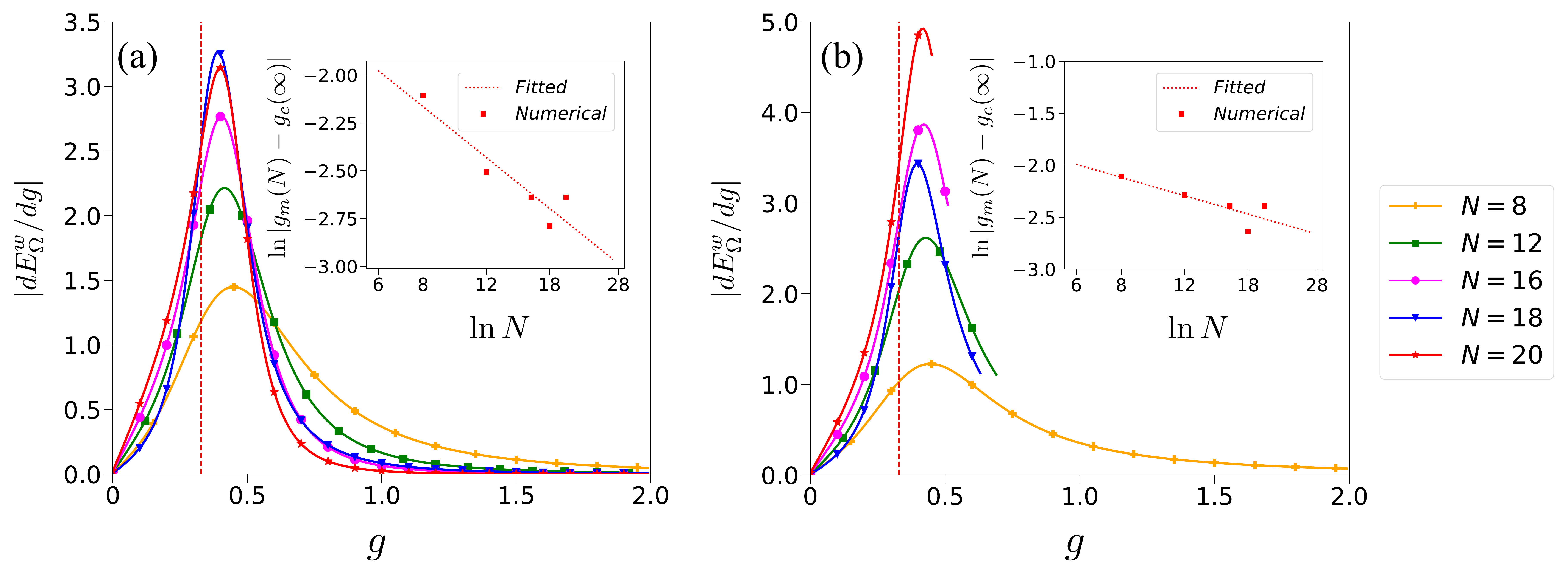}
    \caption{(Color online) \textbf{QPT using witness-based lower bound in Kitaev code.} Variations of $\left|\frac{dE^w_\Omega}{dg}\right|$ as a function of $g$ across the QPT point of a toric code under parallel magnetic field for different system sizes $N$, when $\Omega$ is chosen to be a nontrivial loop representing (a) $L^x_h$ and (b) $L^z_h$. The toric code is considered over a lattice of $N_P^h\times N_P^v$ architecture (see Section~\ref{subsubsec:kitaev}), and the system sizes are taken to be same as in Figure~\ref{fig:fig6}. (Insets of (a) and (b)) The variation of $\ln\left|g_m(N)-g_c(\infty)\right|$ as a function of $\ln N$. The dotted line represents the fitted curve according to Eq.~(\ref{eq:witness_scaling}), with (a)  $\beta=0.44(8)$,$\delta=0.65(5)$ for $\Omega\equiv L^x_h$, and (b) $\beta=0.29(7)$, $\delta=0.43(5)$ for $\Omega\equiv L^z_h$. Note that in (b), $dE^w_\Omega/dg$ data is plotted only till the value of $g$ beyond which $E^w_\Omega=0$ (see Figure~\ref{fig:fig3}). Both the horizontal and the vertical axes in all figures are dimensionless.} 
    \label{fig:fig8}
\end{figure*}

Here, we would like to point out that in the present problem, the nature of the ground state of $H_K$ is qualitatively different in the cases of $g=0$ (degenerate entangled ground state manifold of the topological code) and $g> 0$ (non-degenerate entangled ground state). Therefore, it is reasonable to expect that the preferred set $\mathcal{K}$ corresponding to the set of probabilities $\{p_k\}$ such that $p_k>p_c$ $\forall k\in\mathcal{K}$ is different for $g=0$ and $g>0$. In order to take into account both situations, we consider the preferred set for the entire range of $g$ to be $\mathcal{K}=\mathcal{K}_{0}\cup\mathcal{K}_1\cup\mathcal{K}_2\cup\cdots\cup\mathcal{K}_n$, where $\mathcal{K}_0$ $(\mathcal{K}_i)$ is the preferred set of outcomes corresponding to $g=0$ ($g=g_i>0$, where we choose $n$ non-zero positive values of $g$). The number $n$ and the value of $p_c$ can be judiciously chosen depending on the situation, such that the error $\left|E^{\prime}_\Omega-E^{\prime\prime}_\Omega\right|$ can be minimized depending on the available numerical resource.

To quantitatively estimate the performance of the approximation, let us assume that the set $\mathcal{K}$ contains $K$ elements. Since $0\leq E(\rho^k_\Omega)\leq 1$, it is easy to see that under the approximation in Eq.~(\ref{eq:proposed_lower_bound}), the terms in $E^{\prime\prime}_\Omega$ having value $p_kE(\rho^k_\Omega)<p_c$ are discarded. Therefore the absolute error $\epsilon=\left|E^\prime_\Omega-E^{\prime\prime}_\Omega\right|<\epsilon_m$, where $\epsilon_m =(2^{|\overline{\Omega}|}-K)p_c$ corresponds to each of $2^{|\overline{\Omega}|}-K$ discarded terms in $E^\prime_\Omega$ having the value $p_c$. We choose $p_c=10^{-10}$ and set $n=1$ with $g_1=2\times 10^{-1}$ for all our calculations. This choice ensures $\epsilon_m<10^{-6}$ for all the results presented in this paper even when $g$ has a high value, implying that $E^\prime_\Omega=E^{\prime\prime}_\Omega$ can be assumed in all practical purposes. Figure~\ref{fig:fig5} provides the variation of $\epsilon_m$ as a function of $N$ in the case of nontrivial loops corresponding to $L^x_{h}$ and $L^{z}_{h}$.  Among all instances of the Kitaev code considered in this paper, the maximum value of $\epsilon_m$ occurs for the case of $L^z_{h}$ loop in the case of $N=18$.  Note that even with the use of the approximation introduced above, computation of the entanglement measure $E$ over a $2^{|\Omega|}\times 2^{|\Omega|}$ density matrix $\rho_\Omega^k$ is difficult. In the case of large systems, we use sparse matrix calculations to overcome this hurdle. More specifically, we set density matrix elements  to be zero if its magnitude is $<10^{-8}$.            


\begin{figure*}
    \centering
    \includegraphics[width=0.7\textwidth]{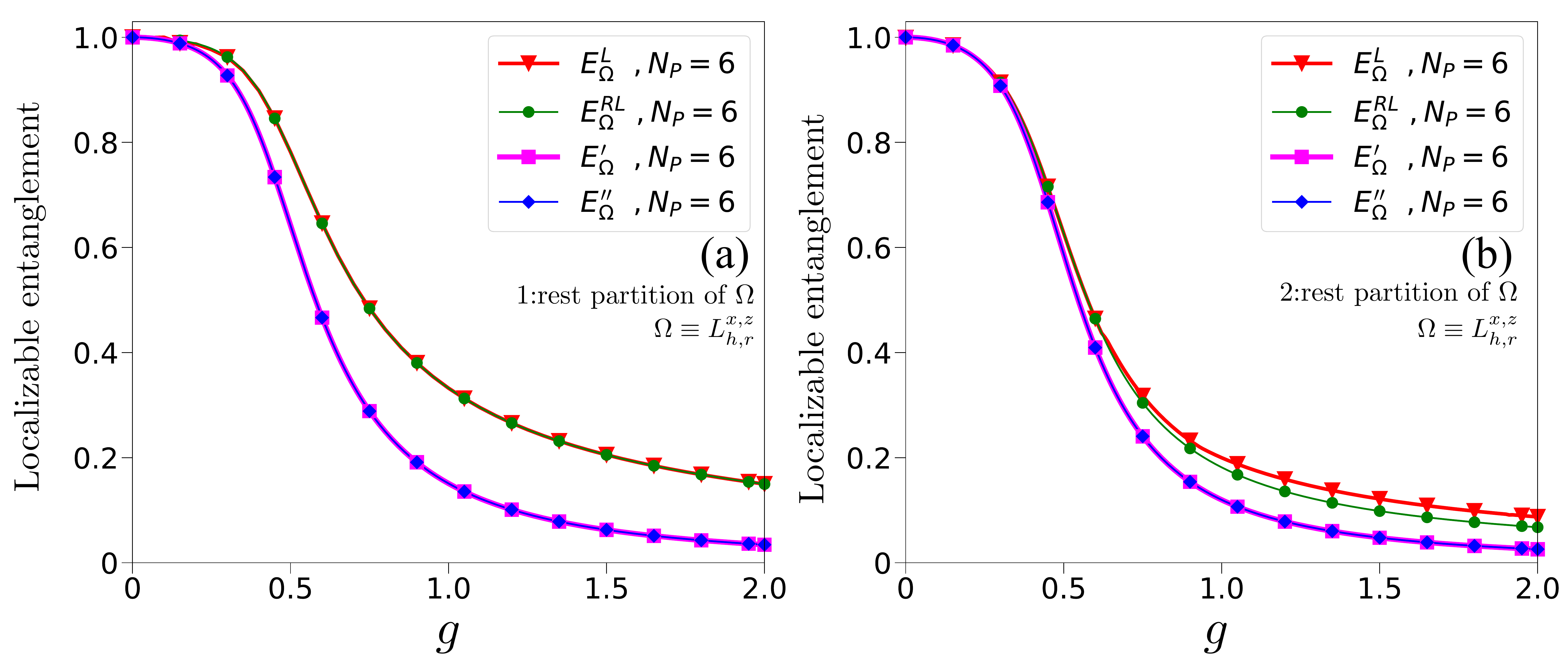}
    \caption{(Color online) \textbf{Lower bounds of localizable entanglement in color codes.} Variations of $E^L_\Omega$, $E^{RL}_\Omega$, $E^{\prime}_\Omega$, and $E^{\prime\prime}_\Omega$ (vertical axes) as functions of the parallel magnetic field strength $g$ (horizontal axes) in the case of a color code of $6$ plaquettes, where entanglement is localized over a nontrivial loop corresponding to $L^x_{h,r}$ in the bipartition (a) $1:\text{rest}$ and (b) $2:\text{rest}$. In the case of localizable entanglement computed over $2:\text{rest}$ partition of the nontrivial loop, we have used normalized negativity as the seed measure for ease of comparison between (a) and (b). Both the horizontal and the vertical axes in all figures are dimensionless.} 
    \label{fig:fig9}
\end{figure*}

\subsubsection{Across the quantum phase transition}

To check whether $E^{\prime\prime}_\Omega$ can signal the topological to nontopological QPT, for each system of size $N$, we focus on $\left|dE^{\prime\prime}_\Omega/dg\right|$ as a function of $g$ (cf.~\cite{banerjee2020} for an investigation of multiparty features of LE), which exhibits a maximum at $g=g_m(N)$ in the vicinity of the QPT point (see Figure~\ref{fig:fig6}) for both cases of entanglement localized over $\Omega\equiv L^x_{h}$ and $L^z_{h}$. Note that the value of $g_m(N)$ is specific to the system size $N$.  With increasing system size, the maximum sharpens and the value $g_{m}(N)$ corresponding to the maximum shifts closer to the QPT point $g_c$, which corresponds to the thermodynamic limit $(N\rightarrow\infty)$. To check how the system approaches the thermodynamic limit, we perform a scaling analysis and investigate the variation of $\ln\left|g_m(N)-g_c(\infty)\right|$ as a function of $\ln N$. We find that the position $g_m(N)$ of the QPT approaches the QPT at thermodynamic limit as (see Figure~\ref{fig:fig6}) 
\begin{eqnarray}
g_m(N)=g_c(\infty)+\alpha N^{-\nu},
\label{eq:le_scaling}
\end{eqnarray}
in the case of $\Omega\equiv L^x_h$, where $\alpha$ is a dimensionless constant, and $\nu$ is the finite-size scaling exponent. For the examples demonstrated in Figure~\ref{fig:fig6}, fitting of numerical data provides $\alpha=0.38(6)$, $\nu=0.58(8)$ for $\Omega\equiv L^x_h$. However, similar analysis with $\Omega\equiv L^z_h$ reveals that while the behavior of $dE^{\prime\prime}/dg$ as a function of $g$ remains the same as in the case of $\Omega\equiv L^x_h$, the finite-size effect is stronger in the former case (see Figure~\ref{fig:fig6}(b)). A finite-size scaling analysis using Eq.~(\ref{eq:le_scaling}) demonstrates a very slow approach of $g_m(N)$ to $g_c$ with increasing $N$, with $\alpha=0.10(2)$, $\nu=0.16(0)$ for $\Omega\equiv L^z_v$. In order to perform a better scaling analysis, one needs to go beyond the system size of $20$ qubit, which, with our available computational resource, remains intractable.  
Note that to perform the scaling analysis, we have considered rectangular lattices ($N=12$ ($3\times 2$), $N=16$ ($4\times 2$), and $N=20$ ($5\times 2$)) along with the square lattices ($N=8$ ($2\times 2$) and $N=18$ ($3\times 3$)), although the topological quantum error corrections are typically performed over Kitaev codes on square lattices~\cite{kitaev2001,kitaev2003}. 

\subsection{QPT from the witness-based lower bound}
\label{subsec:qpt_witness}

In order to show how the topological to nontopological QPT can be accessed experimentally, we adapt the witness-based approach to compute a lower-bound of LE. Following the methodology introduced in ~\cite{amaro2020,alba2010}, a \emph{local} witness operator $W_\Omega$ can be designed such that it's expectation value $\omega=\text{Tr}[W_\Omega\rho]$ in the full state $\rho$ of the system can provide information about the entanglement present in $\Omega$. For topological quantum codes, $W_\Omega$ can be designed exploiting the stabilizers corresponding to the topological codes as~\cite{amaro2018,amaro2020a} 
\begin{eqnarray}
W_\Omega = \frac{I}{2}-\prod_{s_j\in\{s_j\}}\frac{I+s_j}{2},
\label{eq:local_witness}
\end{eqnarray}
where the set $\{s_j\}$ is a subset of $\mathcal{S}$ representing the full set of stabilizers of the topological code under investigation. Each of the elements $s_j$ of the set $\{s_j\}$ can be decomposed as $s_j=s_j^\Omega\otimes s_j^{\overline{\Omega}}$ according to whether the supports of the Pauli matrices constructing $s_j$ belong to $\Omega$ or $\overline{\Omega}$. For $s_j$ to contribute in $W_\Omega$, (a) the Pauli matrices constructing $s_j$ must commute outside $\Omega$, and (b) the set $\{s_j^{\Omega}\}$ obtained from $\{s_j\}$ must be a complete set of stabilizer generators corresponding to a state $\ket{\psi_\Omega}$ over $\Omega$ having genuine multiparty entanglement. Decomposing the witness operators into projection operators over $\overline{\Omega}$ and local witness operators corresponding to the projectors on $\overline{\Omega}$, $w$ can be related to $w_k$ (see Eq.~(\ref{eq:hierarchy3})) as~\cite{amaro2020a} 
\begin{eqnarray}
w=\sum_{k=0}^{2^{|\overline{\Omega}|}-1}p_kw_k.
\label{eq:w_wk_relation}
\end{eqnarray}
Using Eq.~(\ref{eq:w_wk_relation}) and choosing negativity as the bipartite entanglement measure over all $1:\text{rest}$ bipartitions of $\Omega$, $E^w_\Omega$ (Eq.~(\ref{eq:witness_lb})) can be shown to be given by~\cite{amaro2020a} (see also Appendix~\ref{app:witness}) 
\begin{eqnarray}
E^w_{\Omega} &=&
      \begin{cases}
      -2w, & \text{for } w < 0,  \\
      0, & \text{for } w \geq 0.
      \end{cases} 
\end{eqnarray}
Therefore, by computing $\omega$, it is possible to estimate a lower bound of the LE from experiments. Note, however, that the performance of the lower bound depends on the performance of the witness operator. In situations where $w\geq 0$, $E^w_\Omega$ produces the trivial lower bound for LE.

In the case of Kitaev code on rectangular or square lattice, construction of $W_\Omega$ corresponding to $\Omega\equiv L^x_{h}$ ($\Omega\equiv L^z_{h}$) involves a subset of the plaquette (vertex) stabilizers through which $L^x_{h}$ ($L^z_{h}$) passes, and a subset of the vertex (plaquette) stabilizers that share a single qubit with $L^{x}_{h}$ ($L^z_{h}$) (see Figure~\ref{fig:fig7} for a demonstration). The specific forms of the witness operators $W_\Omega$ used for determining $E^w_\Omega$ in the case of $\Omega\equiv L^x_{h}$ and $\Omega\equiv L^z_h$ are shown in Table~\ref{tab:tab1}.  Typical variations of $E^w_\Omega$ obtained from the local witness operators $W_\Omega$ designed for nontrivial loops of Kitaev codes representing $L^x_{h}$ and $L^z_{h}$ as functions of $g$ are shown in Figure~\ref{fig:fig3}.

It is worth pointing out here that the local witness operators are designed for the ground states of the topological quantum codes at $g=0$. In the case of $g>0$, the nature of the ground state of the model changes, and the witness operators $W_\Omega$ may not be able to detect entanglement over $\Omega$, and consequently may fail to reliable provide a lower bound of the LE. Among the witness operators $W_\Omega$ constructed by us for toric codes of different sizes, those designed for the nontrivial loops representing $L^x_{h}$ successfully provide a lower bound of LE for $g>0$, even when $g$ is large. While $E^{\prime\prime}_\Omega\approx E^w_\Omega$ for lower system sizes (see Figure~\ref{fig:fig3}(a) and (b)), the difference between these two bounds increases with increasing $N$ (see Figure~\ref{fig:fig3}(c)). On the other hand, in the case of $L^z_{h}$, expectation value of $W_\Omega$ is found to be positive for higher values of $g$, thereby failing to provide the lower bound once the parallel field is past a critical strength. However, similar to $E^{\prime\prime}_\Omega$, $dE^w_\Omega/dg$ signals the topological to nontopological QPT in both cases of $\Omega\equiv L^x_{h}$ and $\Omega\equiv L^z_{h}$, with a finite size scaling given by
\begin{eqnarray}
g_m(N)=g_c(\infty)+\beta N^{-\delta}.
\label{eq:witness_scaling}
\end{eqnarray}
Here, the dimensionless constant $\beta$ and the finite size scaling exponent $\delta$ have similar significance as $\alpha$ and $\nu$ respectively, which can be determined in a similar fashion as in the case of $E^{\prime\prime}_\Omega$. For example, in the case of a nontrivial loop representing to $L^x_{h}$ (see Figure~\ref{fig:fig8}), $\beta=0.44(8)$ and $\delta=0.65(5)$, while for a nontrivial loop representing $L^z_h$, $\beta=0.29(7)$ and $\delta=0.43(5)$. Note here that although $E^w_{\Omega}=0$ past a certain strength of $g$ for each $N>8$ in the case of $\Omega\equiv L^z_h$, the approach of the QPT point to $g_c$ as demonstrated by $dE^w_\Omega/dg$ is faster compared to that for $dE^{\prime\prime}_\Omega/dg$, as clearly shown by the values of $\delta$ compared to the values of $\nu$ (see Figure~\ref{fig:fig6}(b)).    

\begin{figure*}
    \centering
    \includegraphics[width=0.8\textwidth]{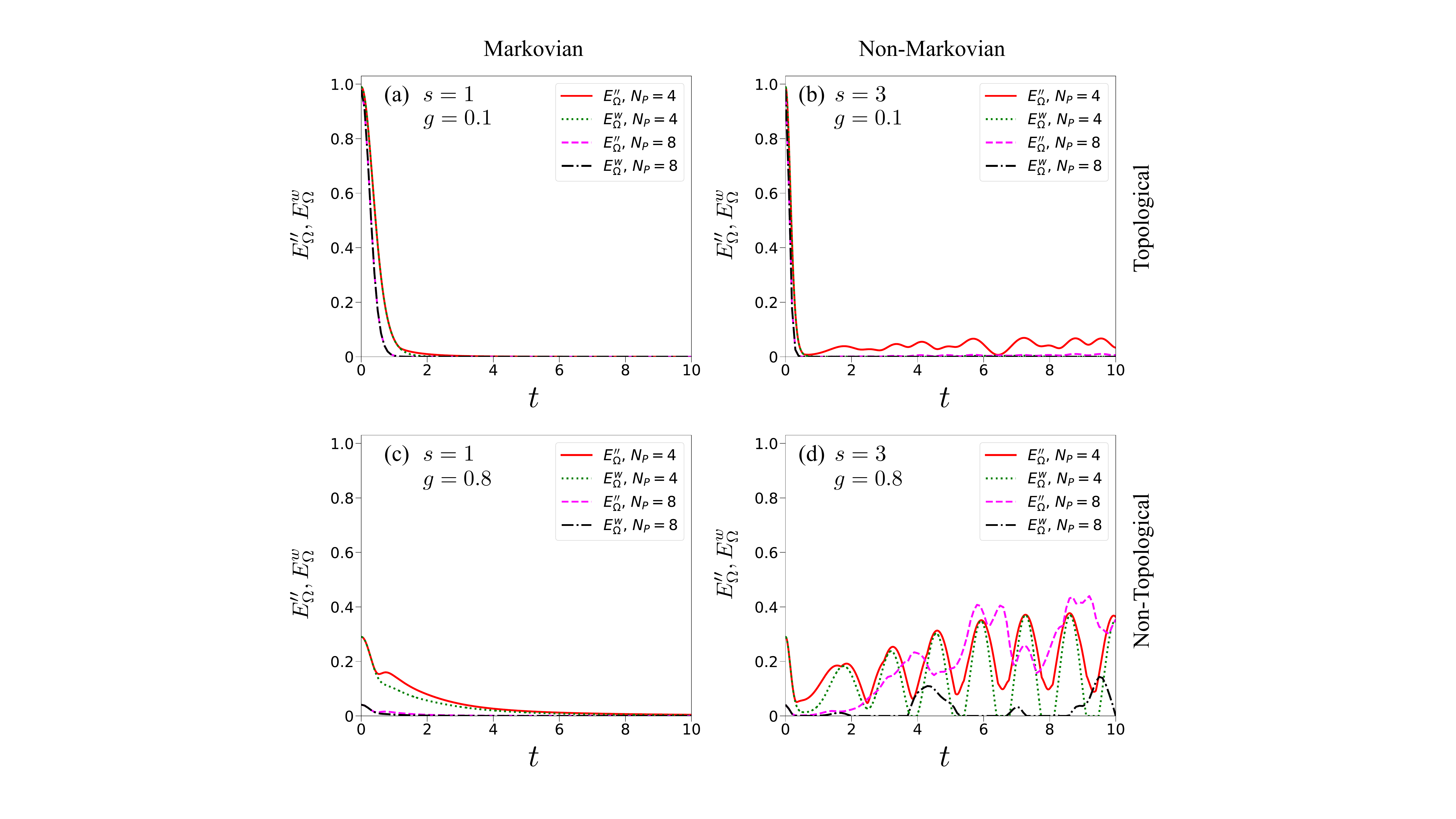}
    \caption{(Color online) \textbf{Markovian and non-Markovian dynamics of localizable entanglement in Kitaev code.} Variations of $E^{\prime\prime}_\Omega$ and $E^w_\Omega$ as functions of $t$ in the case of Markovian ($s=1$) and non-Markovian ($s=3$) dephasing noise, when the initial states of the dynamics are taken from the topological ((a) and (b), $g=10^{-1}$) and nontopological ((c)-(d), $g=8\times 10^{-1}$) phases of the Kitaev code in the presence of parallel magnetic field. The Kitaev model is defined on four- ($N_P=4, 2\times 2$) and eight-plaquette ($N_P=8, 4\times 2$) lattice, and $E^{\prime\prime}_\Omega$ and $E^w_\Omega$ are computed over a nontrivial loop representing $L^x_h$. Both the horizontal and the vertical axes in all figures are dimensionless.}
    \label{fig:fig10}
\end{figure*}


\textbf{Note on the color code.} We also test the approximated lower bound $E^{\prime\prime}_\Omega$ of LE for the color code in a parallel magnetic field (Eq.~(\ref{eq:color_code_hamiltonian})) on a hexagonal lattice. Note that compared to the Kitaev code, the number of qubits in the color code grows faster with the growth of the lattice, thereby making the computation of LE and its bounds more computationally demanding. For demonstration, we focus on the nontrivial loop representing $L^x_{h,r}$ on a 6-plaquette ($12$ qubits) color code (see Figure~\ref{fig:fig1}). We consider a canonical measurement setup such that 
\begin{enumerate}
    \item[(a)] all qubits connected directly to the nontrivial loop with a lattice edge are measured in the $\sigma^x$ basis, and 
    \item[(b)] the rest of the qubits in $\overline{\Omega}$ are measured in the $\sigma^z$ basis,  
\end{enumerate}
which is an extension of the rules for choosing the canonical measurement setup for the Kitaev code. Figure~\ref{fig:fig9} depicts the variations of $E^{L}_{\Omega}$, $E^{RL}_{\Omega}$, $E^{\prime}_{\Omega}$, and $E^{\prime\prime}_\Omega$ as functions of $g$, when the entanglement is localized over a partition of $1:\text{rest}$ and $2:\text{rest}$ of the qubits in $\Omega$. 
The variations of LE and its lower bounds are qualitatively similar to that demonstrated in the case of the Kitaev code. Note, however, that in the case of the color code, equality between $E^{L}_\Omega$ and $E^{RL}_\Omega$ is sustained with an increase in the external field strength only when entanglement is localized over 1:rest partitions of the nontrivial loop. When entanglement is localized over 2:rest partitions, this equality cease to exist at high field value. For computing $E^{\prime\prime}_{\Omega}$, we set values of $n$, $g_i$, and $p_c$ similar to the case of the Kitaev model. Our data indicates that similar to the Kitaev code, $E^{\prime\prime}_\Omega$ provides a good approximation of $E^{\prime}_\Omega$ in the case of the color code also, throughout the range $[0,2.0]$ of $g$. Note that in the case of entanglement localized over $2:\text{rest}$ partition of $\Omega$, the optimization of LE is not covered by the Pauli measurement setup when $g$ is large, as indicated from the deviation of the $E^{RL}_\Omega$ curve from the same of $E^L_\Omega$. 

\section{Localizable entanglement under dephasing noise}
\label{sec:dynamics}

In this section, we discuss the effect of single qubit dephasing noise on the bipartite localizable entanglement over a nontrivial loop of the topological code in the presence of parallel magnetic field. We also demonstrate how the dynamics of localizable entanglement can be used to differentiate between the topological and nontopological phases of the models. We determine the time-dependent state $\rho$ of the system by numerically solving the quantum master equation (Eq.~(\ref{eq:qme})) using the Runge-Kutta $4$th order method with the ground state of the system as the initial state $(t=0)$ $\rho_0$, and then compute the localizable entanglement and its lower bounds as a function of $t$ on a set $\Omega$ of qubits forming a nontrivial loop. In the case of large systems, we use sparse matrix calculations to determine $\rho$ as a function of $t$, setting density matrix elements  to be zero if its magnitude is $<10^{-8}$, as in the case of noiseless situation.

\begin{figure}
    \centering
    \includegraphics[width=0.8\linewidth]{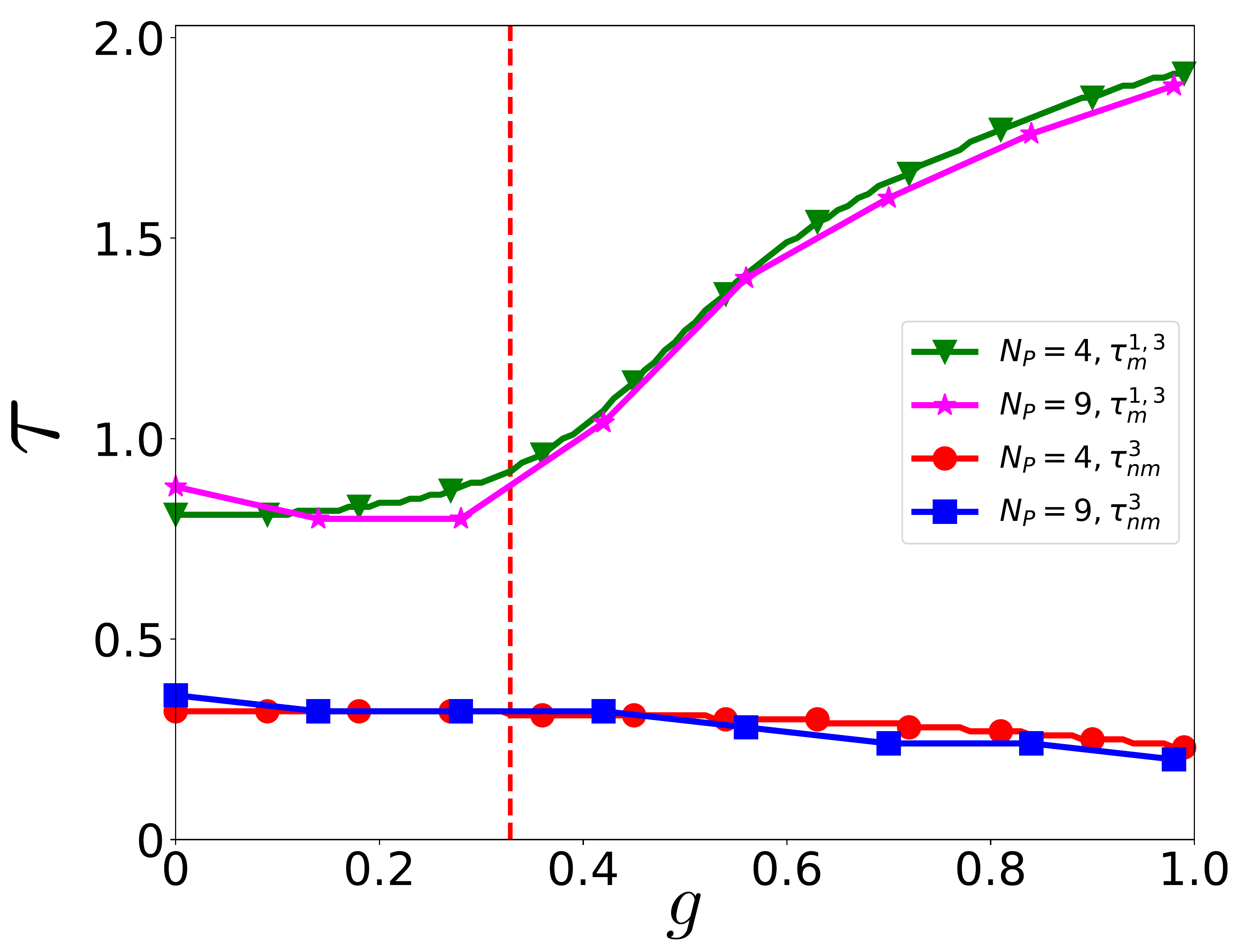}
    \caption{(Color online) \textbf{Entanglement collapse time.} The variations of ECTs, $\tau_{nm}^3$ and $\tau_m^{1,3}$, as functions of $g$ across the QPT point of the Kitaev model in a parallel magnetic field on  $2\times 2$ and $3\times 3$ square lattices, with $\Omega\equiv L^z_h$. The value of $E_c^3$ for computing the ECTs for the Markovian and the non-Markovian dephasing noise is $E_c^3=0.14$ for the $2\times 2$ square lattice, and $E_c^3=0.11$ for the $3\times 3$ square lattice. Both axes are dimensionless. }
    \label{fig:fig11}
\end{figure}

To demonstrate the results, we use the Kitaev model in a parallel magnetic field under the Markovian $(s<2)$ and non-Markovian $(s>2)$ dephasing noise, where we choose the initial state $\rho_0$ to be the ground state of the system in the topological phase $(g<g_c)$, denoted by $\rho_0(g<g_c)$, and the same in the nontopological phase $(g>g_c)$, denoted by $\rho_0(g>g_c)$. We point out that the canonical measurement setup and the local witness operators are designed at the $g=0$ limit of the system Hamiltonian $H_K$, and are demonstrated to work reasonably well (see Section~\ref{sec:optimal_basis}) in the presence of the external magnetic field $g$. However, it is not at all straightforward to have an intuition on how the lower bounds of localizable entanglement devised based on these constructions would perform in the noisy scenarios. To investigate this, we test the hierarchy of the lower bounds of LE in systems of small sizes ($N=8,12$) to find, similar to the noiseless scenario,  
\begin{eqnarray}
E^{L}_\Omega=E^{RL}_\Omega\geq E^{\prime}_\Omega
\end{eqnarray}
for full ranges of $t$ considered in this study, for both cases of $\rho_0(g<g_c)$ and $\rho_0(g>g_c)$. However, as in the case of the noiseless scenario, computation of even  $E^\prime_\Omega$ quickly becomes difficult with increasing lattice size. We further compute $E^{\prime\prime}_{\Omega}$ by setting the values of $n$, $g_1$, and $p_c$ to be the same as in the case of $g=0$ (see Section~\ref{subsubsec:rolb}) to find that for $N=8, 12$, $E^{\prime}_\Omega=E^{\prime\prime}_\Omega$ for all the instances of $\rho_0$ across the QPT point in the case of both Markovian and non-Markovian dephasing noise, and for the full range of $t$. From here onward, we investigate the dynamics of localizable entanglement using $E^{\prime\prime}_\Omega$, and the witness-based lower bound $E^w_\Omega$, which remains computable for a larger system size, and hence can provide an insight into the dynamics of entanglement in large Kitaev codes under parallel magnetic field.

\begin{figure*}
    \centering
    \includegraphics[width=0.7\textwidth]{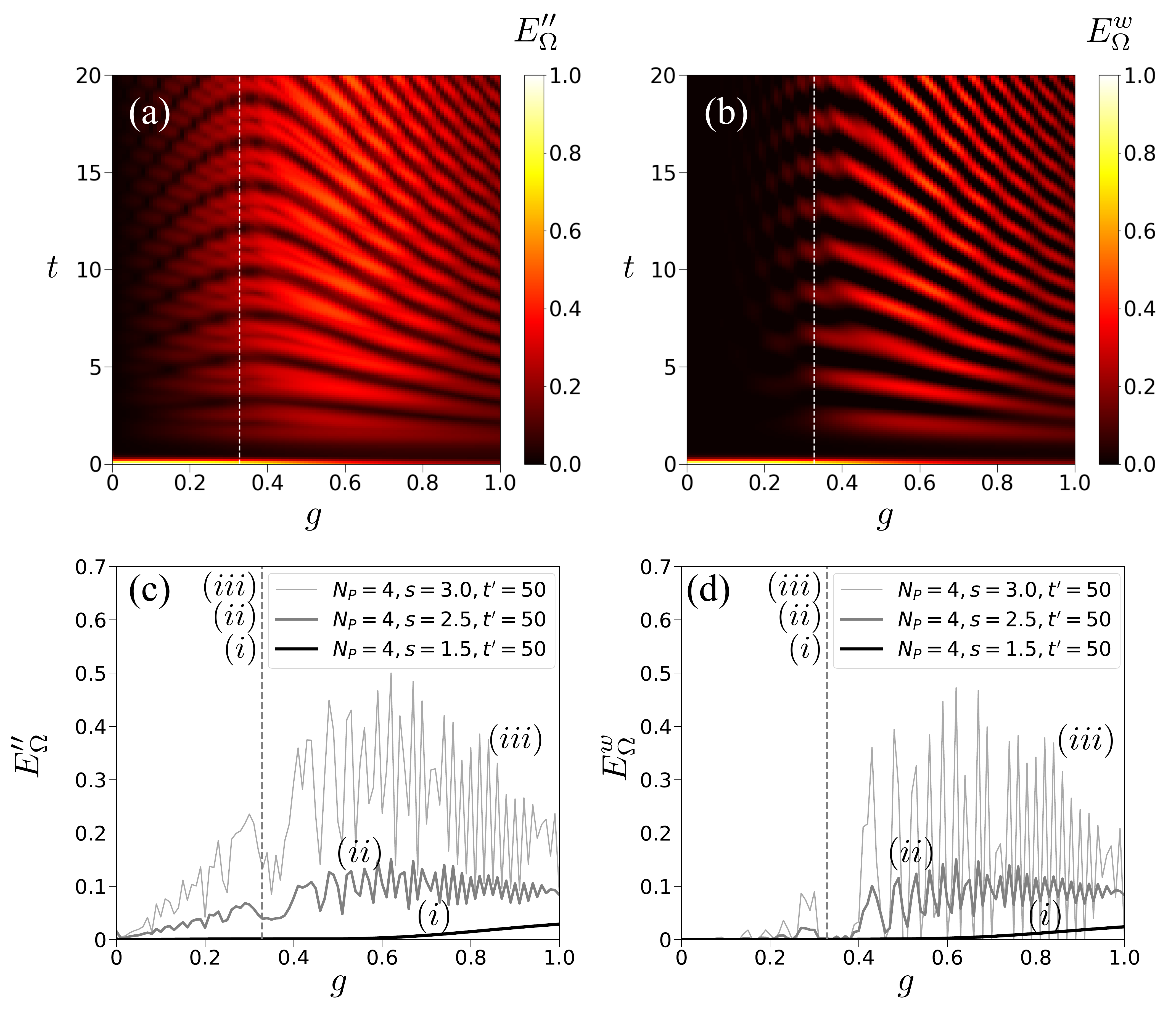}
    \caption{(Color online) \textbf{Distinguishing topological phase from the nontopological phase via dynamics.} Variation of (a) $E^{\prime\prime}_\Omega$ and $E^{w}_\Omega$ as functions of $g$ and $t$. The behaviors of  $E^{\prime\prime}_\Omega$ and $E^w_\Omega$ as functions of $g$ across the QPT point, for fixed values of $t$, are demonstrated in (c) and (d), respectively. All the axes in all figures are dimensionless.} 
    \label{fig:fig12}
\end{figure*}

Typical dynamical behavior of $E^{\prime\prime}_\Omega$ and $E^w_\Omega$ as functions of $t$ are demonstrated in Figure~\ref{fig:fig10} in the case of Markovian and non-Markovian dephasing noise, with initial states chosen from the topological and nontopological phases. In the cases of both initial states $\rho_0(g<g_c)$ (Figure~\ref{fig:fig10}(a)) and $\rho_0(g>g_c)$ (Figure~\ref{fig:fig10}(c)), $E^{\prime\prime}_\Omega$ and $E^w_\Omega$ decay with time when the noise is Markovian ($s=1$), and vanishes at large $t$. Also, no revival of either of $E^{\prime\prime}_\Omega$ and $E^w_\Omega$ is observed at large $t$, which is consistent with the findings on the behavior of entanglement under Markovian noise reported in literature~\cite{yu2009}.  However, under non-Markovian noise $(s=3)$, $E^{\prime\prime}_\Omega$ decays monotonically at first, and then start oscillating with time -- the amplitude of oscillation being much larger in the case of an initial state chosen from the nontopological phase compared to the same for a topological phase (Figure~\ref{fig:fig10}(b), (d)). On the other hand, with initial states chosen from the topological phase, $E^w_\Omega$ monotonically decreases with $t$, goes to zero, and does not revive even at large $t$. In contrast, with initial state chosen from the nontopological phase, $E^w_\Omega$  exhibits large-amplitude oscillations similar to $E^{\prime\prime}_\Omega$, the amplitude increasing with increasing $s$. For high value of the Ohmicity parameter, the dynamical feature of $E^w_\Omega$ resembles  repetitive collapses followed by revivals as $t$ increases, where $E^w_\Omega$ achieves a considerably high value during the revivals.


\paragraph*{Entanglement collapse time.} It is clear from Figure~\ref{fig:fig10} that  in the case of the non-Marokovian noise, the LE and its lower bounds decreases monotonically to a non-zero value $E_{c}^s$ at first, and then starts oscillating. Here, the superscript in $E_c^s$ indicates the value of the Ohmicity parameter, $s$, implying that this value may change with a change in $s$. The \emph{entanglement collapse time} (ECT), for a fixed value of $s>2$, is defined as the time $t=\tau^s_{nm}$ at which the value of $E^{\prime\prime}_\Omega$ collapses to its respective $E_c^s$ value for the first time, such that the landscape of $E^{\prime\prime}_\Omega$ has its first \emph{trough} at $t=\tau^s_{nm}$. For a fixed value of $s$, the value of $\tau^s_{nm}$ can be obtained as the solution of 
\begin{eqnarray}
E^{\prime\prime}_\Omega(t)=E_c^s.
\end{eqnarray}
Similar equations can be written for the LE and all of its lower bounds. In situations where the value of $E^{\prime\prime}_\Omega$ remains at $E_c^s$ for a finite interval of time, $\tau^s_{nm}$ is defined as the first time instant when $E^{\prime\prime}_\Omega(t)=E_c^s$. 

In contrast, in the case of the Markovian noise, $E^{\prime\prime}_\Omega$ decreases monotonically with time, and eventually vanishes. In order to compare the collapse of $E^{\prime\prime}_\Omega$ with the case of non-Markovian noise denoted by a fixed value of $s>2$, in the case of Markovian noise denoted by an Ohmicity $s^\prime<2$, we determine the ECT, $\tau^{s^\prime,s}_{m}$, as the time at which the value of $E^{\prime\prime}_\Omega$, for the first time, collapses to $E_c^s$ corresponding to the non-Markovian noise characterised by $s$. Therefore, $\tau^{s^\prime,s}_m$ can be obtained as the solution of $E^{\prime\prime}_{\Omega}(t)=E_c^s$, where $E^{\prime\prime}_\Omega(t)$ corresponds to the case of the Markovian noise with Ohmicity $s^\prime$. In Figure~\ref{fig:fig11}, we plot the variations of $\tau_{nm}^s$ and $\tau_m^{s^\prime,s}$ as functions of $g$, where $E_c^s$ is chosen depending on the dynamics of $E^{\prime\prime}_\Omega$, in the case of $N_P=4,9$ for the Kitaev code. It is interesting to note that $\tau^s_{nm}$ varies slowly with $g$ across the QPT point, with an overall decreasing trend with increasing $g$. In contrast, in the case of Markovian noise,  $\tau_{nm}^{s^\prime,s}$ remains almost constant in the topological phase, and increases with $g$ in the nontopological phase. Note, however, that by definition,  $\tau_{m}^{s^\prime,s}$ corresponding to the Markovian noise depends on the choice of $E_C^s$ for the non-Markovian noise, which, in turn, depends on the system size as well as the partition over which the localizable entanglement is computed. 

These results indicate that knowledge about the phases of the Kitaev model in a parallel magnetic field can be used to distinguish between the Markovian and non-Markovian type of the dephasing noise using the dynamics of $E^{\prime\prime}_\Omega$ and $E^w_\Omega$. Starting from a ground state of the model in the nontopological phase as the initial state, at $t=t^\prime$, if a highly oscillating behavior with high amplitude of $E^{\prime\prime}_{\Omega}$ is found, then the noise is expected to be non-Markovian. Experimental determination of the type of the noise is also possible via determining $E^w_\Omega$, where non-Markovianity of the dephasing noise is indicated by a repeated revival and collapse of $E^w_\Omega$ over time, when the initial state of the system is chosen from the nontopological phase. Such a distinction can be useful in situations where the single qubit noise is known to be dephasing, but the Markovianity of the noise is not decided.

\subsection{Distinguishing phases from the dynamics}
\label{subsec:distinction}

It is clear from Figs.~\ref{fig:fig10} and the discussion above that in the case of the non-Markovian dephasing noise with high Ohmicity parameter, the magnitudes of oscillations corresponding to $E^{\prime\prime}_\Omega$ are considerably higher, and $E^w_\Omega$ exhibits repeated revivals to high value and then collapses, when the initial state of the dynamics is chosen from the nontopological phase, compared to the same when the initial state of the system is taken from the topological phase. This pattern remains unchanged for considerably wide range of $g$ around the QPT point $g=g_c$, and at large $t$ (see Figure~\ref{fig:fig12}(a) and (b)). Since the initial state ($t=0$) of the dynamics is characteristic to the phases of the system, these feature can be used to distinguish between the phases of the Kitaev code in the presence of parallel magnetic field. Such a distinction is useful in situations where single-qubit dephasing noise is present in the system, and one is forced to investigate the phases after a considerably amount of time has elapsed. The important points of the phase discrimination in this method are as follows.
\begin{enumerate}
\item[(a)] Let us choose a large time $t=t^\prime$, at which $E^{\prime\prime}_{\Omega}(t^\prime)$ and $E^w_\Omega$ is computed. The crossover from the topological to the nontopological phase of the system at $g=g_c$ is distinguished  by (1) the onset of oscillations of large amplitudes of $E^{\prime\prime}_\Omega$, and (2) by the appearance  of repeated revivals and collapses of $E^w_\Omega$, as depicted in Figure~\ref{fig:fig12}(c)-(d) in the case of a four-plaquette system. As the value of the Ohmicity increases, the amplitude of oscillations  for $E^{\prime\prime}_\Omega$ and the maximum value attained by $E^w_\Omega$ during a revival become larger. 

\item[(b)] The large time $t^\prime$ can be chosen according to the situation in hand. In Figure~\ref{fig:fig12}(c)-(d), we have demonstrated $E^{\prime\prime}_\Omega(t^\prime)$ vs. $g$ variations for $t^\prime=50$, such that in the nontopological phase, $t^\prime\gg \tau_{nm}^s$. Our numerical analysis suggests that the oscillations are larger for moderately low values of $g$ in the nontopological phase of the system, while the oscillation gradually dies out as the initial state approaches towards a product state by increasing the value of $g$. Our numerical analysis of the $9$-plaquette system indicates that the qualitative trend of $E^{\prime\prime}_{\Omega}$ and $E^w_\Omega$ vs. $g$ and $t$ remains the same as one goes higher in the system size.
\end{enumerate}

From the above observations, it is clear that the behavior of $E^{\prime\prime}_\Omega$ and $E^w_\Omega$ under non-Markovian dephasing noise can be utilized to distinguish between the topological and the nontopological phase. Note that while a sharp determination of the QPT point is not possible from these features, the phase can be recognized as a topological, or a nontopological one, particularly in experimental scenarios using the witness operators.  

\begin{figure}
    \centering
    \includegraphics[width=0.8\linewidth]{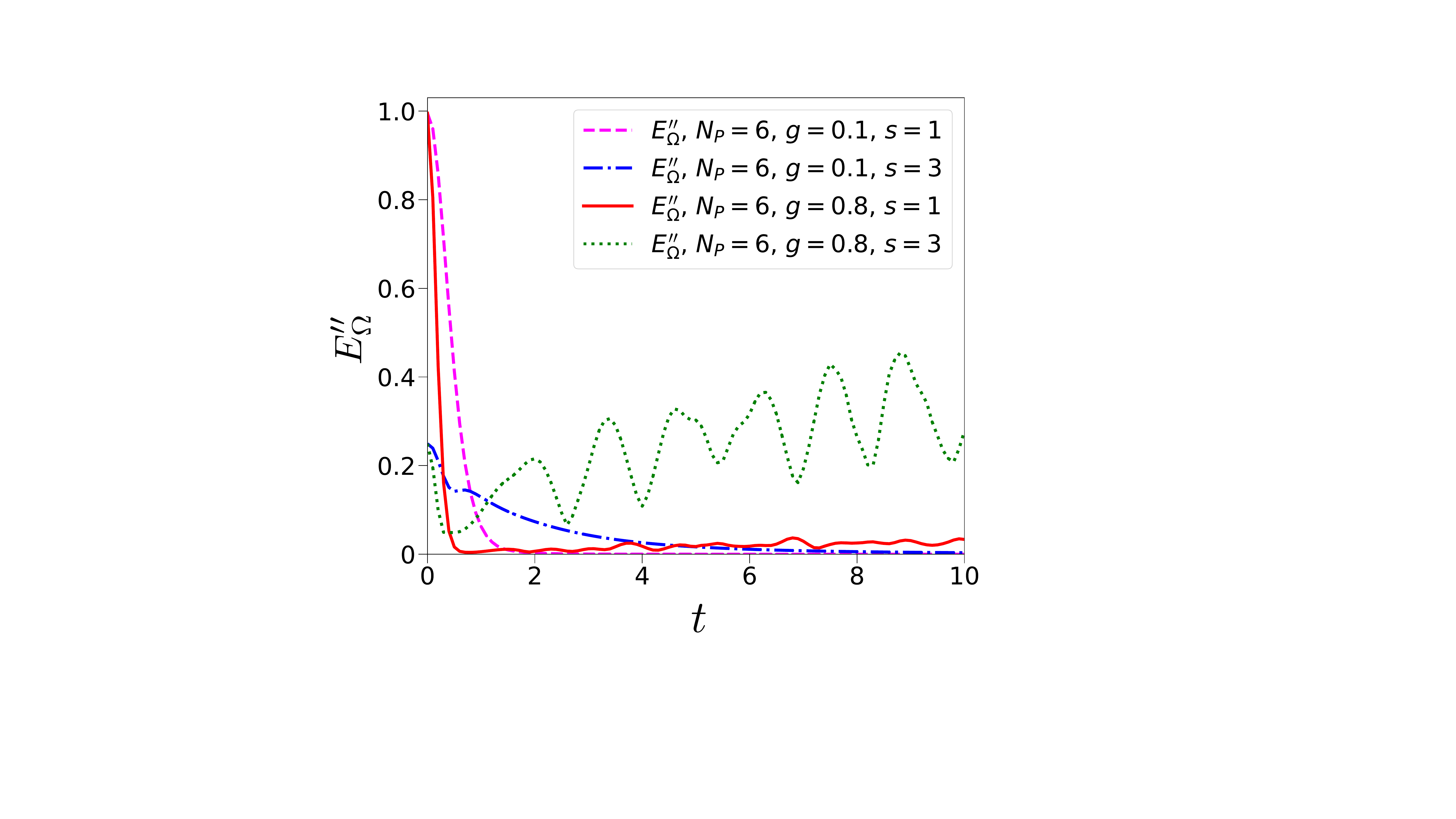}
    \caption{(Color online) \textbf{Markovian and non-Markovian dynamics of localizable entanglement in color code.} Variation of $E^{\prime\prime}_\Omega$ as functions of $t$ in the case of a color code on a hexagonal lattice of $6$ plaquettes in parallel magnetic field, where the initial states of the dynamics are chosen from the topological and the nontopological phases. Both axes are dimensionless.}
    \label{fig:fig13}
\end{figure}

\textbf{Note on the dynamics of color code.} In order to examine whether the features reported above are model-specific, we also test these findings in the case of the color code in a parallel magnetic field on a hexagonal lattice of $6$ plaquettes under periodic boundary condition. We demonstrate the dynamics of $E^{\prime\prime}_\Omega$ in Figs.~\ref{fig:fig13} under the Markovian ($s=1$) and non-Markovian ($s=3$) dephasing noise when the initial state is chosen from the topological and the nontopological phases of the model. We compute $E^{\prime\prime}_\Omega$ using the canonical measurement setup for the color code (see Section~\ref{sec:optimal_basis}), setting the values of $n$, $g_1$, and $p_c$ similar to that reported in Section~\ref{sec:optimal_basis}. The qualitative behaviors of $E^{\prime\prime}_\Omega$ vs. $t$ remains the same as the Kitaev model under parallel magnetic field, as is evident from Figs.~\ref{fig:fig10} and \ref{fig:fig13}, implying that the topological and the nontopological phases in also the color code under parallel field can be distinguished using the dynamics of $E^{\prime\prime}_\Omega$. We point out here that the computation of $E^{\prime\prime}_\Omega$ becomes increasingly difficult in the case of the color code subjected to noise as the system size increases, and therefore a full numerical investigation of the phases of the system via its dynamics is not possible with the available numerical resources.

\section{Conclusion and Outlook}
\label{sec:conclusion}

Topological quantum codes, such as the Kitaev code and the color code, have attracted a lot of attentions due to their immense potential in performing quantum computation tasks. In the presence of external perturbations like local magnetic field, these models exhibit topological order at the zero field limit, which is robust against small perturbations. However, with the increase of the perturbation strength, the model undergoes a topological to nontopological quantum phase transition, and the ground state becomes fully polarized when the field strength is infinite. Such quantum phase transitions are beyond the Landau description of quantum phase transitions, and cannot be probed using the local order parameters and spontaneous symmetry breaking. In this paper, we investigate the topological to nontopological quantum phase transition occurring in a topological quantum code in the presence of a parallel magnetic field in terms of the entanglement localized over a nontrivial loop via local projection measurements on spins outside the loop. To overcome the barrier due to the quantity being computationally demanding, we compute lower bounds of the quantity in terms of a chosen canonical measurement setup, and an appropriately designed witness operator. We also discuss how the phases of the system can be distinguished by observing the dynamical features of these lower bounds when single-qubit dephasing noise is present in the system. 

We conclude with a discussion on possible avenues for future research. Note that within the scope of models discussed in this paper, one may also consider external perturbations other than a parallel magnetic field, such as spin-spin interactions~\cite{trebst2007,zarei2015}, increasing the strength of which takes the system through a topological to nontopological quantum phase transition. Moreover, beyond the topological quantum error correcting codes, it would be interesting to see whether localizable entanglement can be used to probe the topological orders in other lattice models, for example, quantum dimer models~\cite{kalmeyer1987,rokhsar1988,read1989,moessner2001}, chiral spin states~\cite{wen1989,wen1990a}, and spin liquid states~\cite{wen1991,wen2002}. Also, in order to be able to quantitatively distinguish the phases of topological quantum error correcting codes under external perturbations in the presence of noise other than the dephasing noise, analysis of the dynamical features of localizable entanglement can be performed in the case of the depolarizing and the amplitude-damping noise~\cite{nielsen2010}, which are among the commonly occurring noises in experiments~\cite{schindler2013,bermudez2017}.      

\acknowledgements 
We acknowledge the support from the Science and Engineering Research Board (SERB), India through the Start-Up Research Grant (SRG) (File No. -  SRG/2020/000468 Date: 11 November 2020), and the use of QIClib (\href{https://github.com/titaschanda/QIClib}{https://github.com/titaschanda/QIClib}) -- a modern C++ library for general purpose quantum information processing and quantum computing. We also thank the anonymous Referees for valuable suggestions. AKP thanks Aditi Sen(De) for useful discussions on non-Markovian noise.

\appendix

\section{Negativity as a bipartite entanglement measure}
\label{app:negativity}

Negativity for a bipartite quantum state $\rho_{AB}$ is defined as
\begin{eqnarray} 
\mathcal{N}(\rho_{AB})=||\rho_{AB}^{T_{A}}||-1,
\label{eq:neg_def}
\end{eqnarray}
with $||\varrho|| = \mbox{Tr}\sqrt{\varrho^{\dagger}\varrho}$ being the trace norm of the density operator $\varrho$, and $\rho_{AB}^{T_{A}}$ is obtained by performing partial transposition of $\rho_{AB}$ w.r.t. the party $A$~\cite{peres1996,horodecki1996,zyczkowski1998,lee2000,vidal2002,plenio2005}. The normalized negativity is given by $\mathcal{N}(\rho_{AB})=\left[||\rho_{AB}^{T_{A}}||-1\right]/(d-1)$~\cite{leggio2020}, $d$ being the minimum of the dimensions of the Hilbert spaces of the subsystems $A$ and $B$, such that $0\leq\mathcal{N}\leq 1$. Unless otherwise sated, we always use Eq.~(\ref{eq:neg_def}) to compute negativity. 

\begin{figure*}
\includegraphics[width=0.8\textwidth]{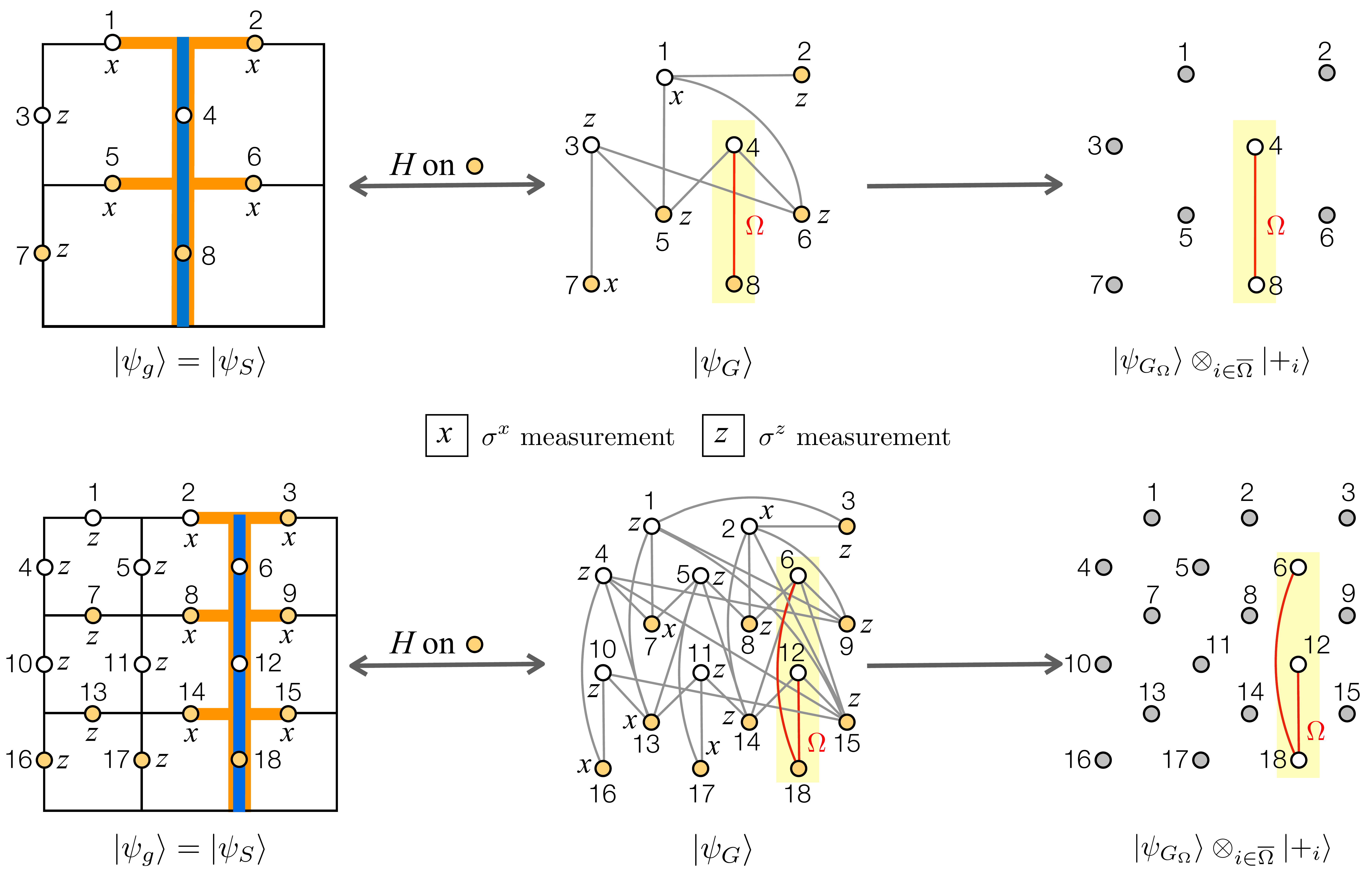}
\caption{(Color online) \textbf{Examples of canonical measurement setups for Kitaev code.} In the case of the $4$-plaquette Kitaev code, the local unitary equivalent graph state $\ket{\psi_G}$ can be obtained by application of the Hadamard operator on a set of control qubits. The corresponding graph $G$ has a connected subgraph $G_\Omega$ over $\Omega$, made of qubits $4$ and $8$ (i.e., the subsystem $\Omega\equiv L^z_v$) connected by a link. An optimal measurement setup that does not disturb $G_\Omega$ during the measurement corresponds to $\sigma^x$ measurements on qubits $1,7$, and $\sigma^z$ measurements on the rest. This is equivalent to $\sigma^z$ measurements on qubits $3,7$, and $\sigma^x$ measurements on  the rest in the ground state $\ket{\psi_g}$ of the Kitaev code, thereby constructing the canonical measurement setup described in Section~\ref{subsec:canonical} (see also Figure~\ref{fig:fig4}). The same for a $9$-plaquette Kitaev code has also been demonstrated, leading to a connected subgraph on qubits $6$, $12$, and $18$ (constructing $L^z_v$) after measurement. Note that in $G_\Omega$ corresponding to the $9$-plaquette Kitaev code, qubit $18$ serves as the hub. Once measurements are made on the graph state, the post-measures states of the system is $\ket{\psi_{G_\Omega}}\otimes_{i\in\overline{\Omega}}\ket{+_i}$ up to local unitary operations, where $\ket{\psi_{G_\Omega}}$ is the state corresponding to the star graph on $\Omega$.}
\label{fig:fig14}
\end{figure*}

\section{Canonical measurement setup for topological quantum codes}
\label{app:canonical_setup}

Here we discuss the canonical measurement setup for computing the localizable entanglement over the group of qubits constituting a nontrivial loop in a topological quantum code.

A graph state~\cite{hein2004,hein2006} is a genuinely multiparty entangled quantum state defined over a simple, connected, and undirected graph $G(V,L)$ made of a collection of nodes, $V$, and links, $L$, as
\begin{eqnarray}
\ket{\psi_G}&=&\left[\otimes_{(i,j)\in L}C_{ij}^z\right]\ket{+}^{\otimes N}, 
\label{eq:graph_state}
\end{eqnarray}
with
\begin{eqnarray}
C^z_{ij}=\frac{1}{2}\left[(I_{i}+\sigma^z_{i})I_{j}+(I_{i}-\sigma^z_{i})\sigma^z_{j}\right]
\end{eqnarray}
being an entangling controlled phase gate. Here, $\ket{+}=(\ket{0}+\ket{1})/\sqrt{2}$ is the eigenstate of $\sigma^x$ corresponding to the eigenvalue $+1$, $(i,j)\in L$ denotes a link in the graph, and $i=1,2,\cdots,N$ labels the nodes in $V$, each holding a qubit. Each ground state $\ket{\psi_g}$ of $H_K$ or $H_C$ at $g=0$ is a stabilizer state $\ket{\psi_S}$, which can be transformed to a graph state $\ket{\psi_G}$ via local Clifford unitary operators as~\cite{lang2012,amaro2020a}
\begin{eqnarray}
\ket{\psi_G}=U\ket{\psi_S};\quad U=\otimes_{i=1}^N U_i.
\end{eqnarray}
Let us now assume that the optimal measurement over $\overline{\Omega}$ that results in the maximum value of $E^L_\Omega$ over a subset of qubits $\Omega$ in $\ket{\psi_G}$ is given by $M_{\overline{\Omega}}$, and the corresponding post-measured states and probability of outcomes are given by 
\begin{eqnarray}
\rho^k &=& \frac{1}{p_k}\left[\left(M_{\overline{\Omega}}\otimes I_{\Omega}\right)\rho_G \left(M_{\overline{\Omega}}\otimes I_{\Omega}\right)\right]
\end{eqnarray}
and 
\begin{eqnarray}
p_k &=& \text{Tr}\left[\left(M_{\overline{\Omega}}\otimes I_{\Omega}\right)\rho_G \left(M_{\overline{\Omega}}\otimes I_{\Omega}\right)\right]
\end{eqnarray}
respectively, with $\rho_G=\ket{\psi_G}\bra{\psi_G}$. Since $\rho_G=U\rho_S U^\dagger$ with $\rho_S=\ket{\psi_S}\bra{\psi_S}$, $\rho_G$ and $\rho_S$ have identical entanglement properties due to their local unitary connection, the optimal value of $E^L_{\Omega}$ in $\rho_G$ corresponds to the same value of $E^L_\Omega$ in $\rho_S$, but corresponding to a different optimal measurement given by $M^\prime_{\overline{\Omega}}=U_{\overline{\Omega}}M_{\overline{\Omega}}U^\dagger_{\overline{\Omega}}$, where $U_{\overline{\Omega}}=\otimes_{i\in\overline{\Omega}}U_i$, and  $U=U_{\overline{\Omega}}\otimes U_{{\Omega}}$. This can be easily seen as 
\begin{eqnarray}
\rho^k &=& \frac{1}{p_k}\left[U\left(M^\prime_{\overline{\Omega}}\otimes I_{\Omega}\right)\rho_S \left(M^\prime_{\overline{\Omega}}\otimes I_{\Omega}\right)U^\dagger\right].
\end{eqnarray}

Single-qubit Pauli measurements on the qubits in $\overline{\Omega}$ of a graph state can be translated to a set of local graph operations~\cite{hein2006}, leading to a graph state on the unmeasured qubits. In the case of two-qubit subsystem $\Omega$, prescription involving single-qubit Pauli measurements on the qubits in $\overline{\Omega}$ exists~\cite{hein2004,hein2006}, which leads to a connected qubit pair over $\Omega$. This is  equivalent to a Bell state up to local unitary operations, thereby ensuring maximum bipartite entanglement. Since $U$ belongs to the class of single-qubit Clifford operations, it follows straightforwardly that the optimal measurement setup for $E^L_\Omega$ in $\rho_S$ for the two-qubit subsystem $\Omega$ is also constituted of single-qubit Pauli measurements over the qubits in $\overline{\Omega}$. The situation is more complex if $\Omega$ contains more than two qubits, since for bipartite entanglement measures, maximum entanglement over all possible bipartitions of $\Omega$ in the post-measured state $\rho^k$ is not guaranteed by single-qubit Pauli measurements on the qubits in $\overline{\Omega}$~\cite{hein2006}. However, in situations where a Pauli measurement setup results in a connected subgraph over $\Omega$, maximum bipartite entanglement is guaranteed in a bipartition of $\Omega$ as long as one of the partitions is a qubit alone, due to the maximally mixed single-qubit density matrix ensured by a connected graph~\cite{hein2006,verstraete2003}. Therefore, in an approach similar to the two-qubit subsystem, one can determine an optimal measurement setup for $\rho_S$ in terms of local Pauli measurements on the qubits in $\overline{\Omega}$, when the entanglement is localized over a $1:\text{rest}$ bipartition.

In this paper, we focus on localizing entanglement over $1:\text{rest}$ partitions of $\Omega$, which allows us to exploit the connection between the graph states and stabilizer states discussed above in determining a canonical measurement setup. We describe this using the Kitaev code. The Clifford operators connecting a stabilizer ground state of the code to graph states are Hadamard operators given by 
\begin{eqnarray}
H=\frac{1}{\sqrt{2}}\left(\ket{0}\bra{0}+\ket{0}\bra{1}+\ket{1}\bra{0}-\ket{1}\bra{1}\right)
\end{eqnarray}
on selected control qubits on the code, resulting in the transformation $\sigma^x\overset{H}{\longleftrightarrow}\sigma^z$, with $\{\ket{0},\ket{1}\}$ being the single-qubit computational basis. For a specific set $S_c$ of such control qubits, the corresponding graph state $\ket{\psi_G}=\otimes_{i\in S_c} H_i$ can be determined. Any Pauli measurement setup on the qubits in $\overline{\Omega}$ leading to a connected graph over $\Omega$ can potentially be a canonical measurement setup. However, the set $S_c$ can be chosen in such a way that the subgraph $G_\Omega$ over $\Omega$ in $G$ is already connected~\cite{lang2012,HK_inprep}, taking the form of a \emph{star graph}. In a star graph, one node, referred to as the \emph{hub}, is connected to all other nodes in $\Omega$ via a link, and no other nodes are connected to each other. Note that there may exist multiple possibilities for $S_c$ leading to a star graph on $\Omega$, and any Pauli measurement over the qubits in $\overline{\Omega}$ that (a) either does not disturb the subgraph $G_\Omega$, or (b) results in another connected subgraph $G_\Omega$, can be chosen as a canonical measurement setup. We choose the former, where
the measurement over the qubits in $\overline{\Omega}$ can be described via simple rules using the relative positions of the qubits with respect to the loop operator, and the plaquette and the vertex stabilizer operators through which the loop passes, as discussed in Section~\ref{subsec:canonical}. These rules are as follows.
\begin{enumerate}
\item[(a)] To determine the localizable entanglement over a nontrivial loop representing the logical operator $L^x_{h,v}$ ($L^z_{h,v}$), a qubit which is not on $L^x_{h,v}$ ($L^{z}_{h,v}$) but is situated on the plaquette operators (vertex operators) through which $L^x_{h,v}$ ($L^z_{h,v}$) passes, is measured in the $\sigma^z$ ($\sigma^x$) basis. 
\item[(b)] All other qubits are measured in the $\sigma^x$ ($\sigma^z$) basis. 
\end{enumerate}
See Figure~\ref{fig:fig14} for a demonstration on Kitaev code with $\Omega$ corresponding to a nontrivial loop representing $L^z_v$.

Note that a measurement setup obtained following these rules works only in the case of $g=0$, and may fail to provide the optimal $E^{RL}_\Omega$ over $\Omega$ in the presence of perturbations in the form of either external field or environmental noise. Examples of such scenarios are reported in Secs.~\ref{sec:optimal_basis} and \ref{sec:dynamics}.

\section{Lower bound from witness operators}
\label{app:witness}

Here we discuss the computation of the witness-based lower bound for localizable entanglement in topological quantum code. Noticing that $s_j=s_j^{\Omega}\otimes s_j^{\overline{\Omega}}$ (see Section~\ref{subsec:qpt_witness}), local witness operator $W_\Omega$ (Eq.~(\ref{eq:local_witness})) can be decomposed as~\cite{amaro2020a} 
\begin{eqnarray}
W_\Omega = \sum_{k=0}^{2^{|\overline{\Omega}|}-1} P_{k,\alpha}^{\overline{\Omega}}\otimes W_\Omega^k.
\end{eqnarray}
Here, 
\begin{eqnarray}
W_\Omega^k = \frac{I}{2}-\prod_{s_j\in\mathcal{S}}\frac{I_\Omega+\eta_j s_j^\Omega}{2}
\end{eqnarray}
with $\eta_j=\pm1$ and $I$ being the identity operator in the Hilbert space of $\Omega$, and
\begin{eqnarray}
P^{\overline{\Omega}}_{k,\alpha} =\frac{1}{2^{|\overline{\Omega}|}}\otimes_{i\in\overline{\Omega}}[I_i+(-1)^{k_i}\sigma^{\alpha_i}], 
\end{eqnarray}
where $k_i=0,1$, and $\alpha_i=x,y,z$. Without loss of any generality, one may label the qubits in $\overline{\Omega}$ as $1,2,\cdots,|\overline{\Omega}|-1,|\overline{\Omega}|$, such that the indices $k=k_1k_2\cdots k_{|\overline{\Omega}|}$ and $\alpha\equiv \alpha_1\alpha_2\cdots \alpha_{|\overline{\Omega}|}$ can be identified as multi-indices, and each sequence $\alpha$ represents a specific Pauli-measurement setup over $\overline{\Omega}$, with $P^{\overline{\Omega}}_{k,\alpha}$ representing the projection operator corresponding to the measurement outcome $k$ for the Pauli measurement sequence specified by $\alpha$. Therefore, the expectation value of $W_\Omega$ in the stabilizer state $\rho_S=\ket{\psi_S}\bra{\psi_S}$ of the topological quantum code is given by~\cite{amaro2020a} 
\begin{eqnarray}
w&=&\text{Tr}[W_\Omega\rho_S]=\sum_{k=0}^{2^{|\overline{\Omega}|}-1}p_k\text{Tr}[W_\Omega^k\rho_\Omega^k]=\sum_{k=0}^{2^{|\overline{\Omega}|}-1}p_k w_k.\nonumber\\ 
\label{eq:witness_relation}
\end{eqnarray}
Note that given an entanglement measure $E$, we would like to determine $E_{\min}(w)$ (see Eq.~(\ref{eq:witness_lb})). The relation between  $w$ and $w_k$ in Eq.~(\ref{eq:witness_relation}) suggests that $E_{\min}(w)$ provides a lower bound of the LE as long as $E_{\min}(w)$ is linear in $w$, where $E$ is the chosen entanglement measure. 

In order to determine $E_{\min}(w)$, we again exploit the connection between the graph states and the stabilizer states (see Appendix~\ref{app:canonical_setup}), and note that~\cite{amaro2020a}
\begin{eqnarray}
w=\text{Tr}[W_\Omega\rho_S]=\text{Tr}[U W_\Omega U^\dagger\rho_G]=\text{Tr}[W_\Omega^\prime\rho_G],
\end{eqnarray}
where the local unitary transformed witness operators  $W^\prime_\Omega=UW_\Omega U^\dagger$  involve stabilizers $s^\prime_j =Us_jU^\dagger$, which are graph state generators. These graph state generators can also be written as $s_j^\prime=s_j^{\prime\Omega}\otimes s_j^{\prime\overline{\Omega}}$, where $s_j^{\prime\Omega}$ represent the graph state generators corresponding to the subgraph $G_\Omega$ corresponding to the subsystem $\Omega$. Witness operators $\tilde{W}_\Omega^{\prime}$ can further be constructed for the state $\rho_{G_\Omega}$ corresponding to the  subsystem $\Omega$ using the graph state generators $s_j^{\prime\Omega}$, such that~\cite{guhne2005,amaro2020} 
\begin{eqnarray}
    w=\text{Tr}[W^\prime_\Omega\rho_G]=\text{Tr}[\tilde{W}_\Omega^\prime\rho_{G_\Omega}]. 
\end{eqnarray}
Note that $\tilde{W}^\prime_\Omega$ is an witness operator that is global to the state~$\rho_{G_\Omega}$.

For a specific stabilizer state $\rho_S$ corresponding to a topological code transformed to a graph state $\rho_G$ via a specific set of unitary operators $U$, and consequently for a specific $W_\Omega^\prime$ resulting in a specific $\tilde{W}^\prime_\Omega$, the optimization problem becomes $E_{\min}(w)=\inf E(\rho)$ subject to $w=\text{Tr}[\tilde{W}_\Omega^\prime\rho]$, $\rho>0$, and $\text{Tr}[\rho]=1$. We choose negativity as the bipartite measure $E$ of entanglement over a bipartition $A:B$ of $\Omega$. Using the definition (see Appendix~\ref{app:negativity}), determination of $E^w_\Omega$ reduces to the optimization problem~\cite{amaro2018,eisert2007} 
\begin{eqnarray}
E^w_\Omega = \inf||(\rho_{AB})^{T_A}||_1-1
\end{eqnarray}
over all possible bipartite state $\rho_{AB}$ such that $\text{Tr}[\tilde{W}_\Omega^\prime\rho_{AB}]=w$, $\text{Tr}[\rho_{AB}]=1$, and $\rho_{AB}>0$. Following the procedure described in described in~\cite{eisert2007}, this turns out to be 
\begin{equation}
    E_{\Omega}^w= \inf \text{Tr}[(D\rho_{AB})^{T_A}]-1,
\end{equation}
with $D$ being an operator satisfying $||D||_{\infty}=1$. Following~\cite{eisert2007} and assuming
\begin{equation}
    D=-f(\tilde{W}^\prime_\Omega)^{T_A}+hI
\label{eq:D}    
\end{equation}
with the factors $f,h$ suitably chosen to meet the condition $||D||_{\infty}=1$, one obtains
\begin{equation}
     E_{\Omega}^w=\max_{f,h}(-fw+h-1).
     \label{eq:optimizable_function}
\end{equation}

We now apply more restrictions specific to our problem in order to compute $({\tilde{W}}^\prime_{\Omega})^{T_A}$, and write
\begin{eqnarray}
({\tilde{W}}^\prime_{\Omega})^{T_A} = \frac{1}{2}I-\rho_{AB}^{T_A},
\label{eq:witness_transpose}
\end{eqnarray}
where $\rho_{AB}$, in general, is a graph state $\rho_{G_\Omega}$. The partially transposed graph state is diagonal in the graph-state basis~\cite{hein2004,hein2006}, and its form can be derived depending on the structure of the graph $G_\Omega$. As per the discussion in Section~\ref{subsec:canonical} and Appendix~\ref{app:canonical_setup}, we focus on transformations $\rho_S\rightarrow\rho_G$ with the restriction of obtaining a connected star graph $G_\Omega$ over $\Omega$. Moreover, we fix a Pauli measurement setup in the form of the canonical measurement setup that does not disturb $G_\Omega$. Therefore, it is sufficient to focus on star graphs $\rho_{G_\Omega}$ for the optimization in Eq.~(\ref{eq:optimizable_function}). Such a graph state can be partially transposed considering the hub to be the partition $A$ and the rest of the nodes to be partition $B$, as~\cite{hein2004,hein2006},
\small 
\begin{eqnarray}
    \rho_{AB}^{T_A}&=&\frac{1}{2}\big[Z_0\rho_{AB}Z_0+Z_{2^{|\Omega|-1}}\rho_{AB}Z_{2^{|\Omega|-1}}\nonumber\\
    &&+Z_{2^{|\Omega|-1}-1}\rho_{AB}Z_{2^{|\Omega|-1}-1}-Z_{2^{|\Omega|}-1}\rho_{AB}Z_{2^{|\Omega|}-1}\big],\nonumber\\ 
    \label{eq:partial_transpose_formula}
\end{eqnarray}\normalsize 
where $Z_l=\otimes_{i\in\Omega}(\sigma^z_{i})^{l_i}$, $l=0,1,\cdots,2^{|\Omega|}-1$, $l_i=0,1$, $l\equiv l_1l_2\cdots l_{|\Omega|}$ is a multi-index having the decimal values of the binary string $l_1l_2\cdots l_{|\Omega|}$. For example, in the case of a subgraph $G_\Omega$ constituted of three qubits ($|\Omega|=3$), such as the graph obtained from the $9$-plaquette Kitaev code (see Figure~\ref{fig:fig14}), $\rho_{AB}^{T_A}$ is given by 
\small 
\begin{eqnarray}
    \rho_{AB}^{T_A}&=&\frac{1}{2}[Z_0\rho_{AB}Z_0+Z_{4}\rho_{AB}Z_{4}+Z_{3}\rho_{AB}Z_{3}-Z_{7}\rho_{AB}Z_{7}].\nonumber\\
\end{eqnarray}\normalsize 
Note that Eq.~(\ref{eq:partial_transpose_formula}) works only when the subsystem $A$ is taken to be the hub, and therefore is not invariant under qubit permutations within $G_\Omega$. Note also that via local graph transformations that results in local unitary transformations over the graph state $\rho_{G_\Omega}$, the graph $G_\Omega$ can be transformed to star graphs with different qubits as hubs. Using Eqs.~(\ref{eq:D}), (\ref{eq:witness_transpose}), and (\ref{eq:partial_transpose_formula}), one obtains    
\small 
\begin{eqnarray}
D&=&\left(h-\frac{f}{2}\right)I-\frac{f}{2}\big[Z_0\rho_{AB}Z_0+Z_{2^{|\Omega|-1}}\rho_{AB}Z_{2^{|\Omega|-1}}\nonumber\\ 
&&+Z_{2^{|\Omega|-1}-1}\rho_{AB}Z_{2^{|\Omega|-1}-1}-Z_{2^{|\Omega|}-1}\rho_{AB}Z_{2^{|\Omega|}-1}\big].\nonumber\\
\end{eqnarray}\normalsize 
with singular values given by $\{|h|,|h-f|,|h-\frac{f}{2}|\}$.

Since $|h-f|>|h-f/2|$, satisfaction of $||D||_{\infty}=1$ requires  $\max\{|h|,|h-f|\}=1$. Under this condition, the optimum values of $h$ and $f$ that maximizes Eq.~(\ref{eq:optimizable_function}) are (a) $h=1,f=2$ for $w<0$, and (b) $h=1,f=0$ for $w\geq 0$. Substituting for $h$ and $f$ in Eq.~(\ref{eq:optimizable_function}), we obtain 
\begin{eqnarray}
E^w_{\Omega} &=&
      \begin{cases}
      -2w, & \text{for } w < 0,  \\
      0, & \text{for } w \geq 0.
      \end{cases} 
\end{eqnarray}

\bibliography{ref}{}

\end{document}